\documentclass[aps,prd,nofootinbib]{revtex4}
\usepackage[colorlinks=true, pdfstartview=FitV, linkcolor=blue, citecolor=red, urlcolor=magenta]{hyperref}
\usepackage{graphicx}
\usepackage{latexsym}
\usepackage{amsmath}
\usepackage{amsfonts}
\usepackage{amssymb}
\usepackage{verbatim}
\usepackage{dsfont}
\usepackage{pbox}
\usepackage{array}   

\newcommand{\be}{\begin{equation}}
\newcommand{\ee}{\end{equation}}
\newcommand{\bea}{\begin{eqnarray}}
\newcommand{\eea}{\end{eqnarray}}

\newcommand{\ben}{\begin{eqnarray}}
\newcommand{\een}{\end{eqnarray}}

\begin{document}

\title{Boundary effects in classical liquid density fluctuations at finite temperature}


\author{$^{1}$ Herondy Mota}
\email{hmota@fisica.ufpb.br}

\author{$^{1,2}$K. E. L. de Farias}
\email{klecio.lima@academico.ufpb.br}

\affiliation{$^{1}$Departamento de F\' isica, Universidade Federal da Para\' iba,\\  Caixa Postal 5008, Jo\~ ao Pessoa, Para\' iba, Brazil.}
\affiliation{$^{2}$Departamento de F\'{\i}sica, Universidade Federal de Campina Grande,\\
Caixa Postal 10071, 58429-900, Campina Grande, Para\'{\i}ba, Brazil.}


\begin{abstract}
We investigate thermal effects on density fluctuations in confined classical liquids using phonon quantization. The system is modeled via a massless scalar field between perfectly reflecting parallel planes with Dirichlet, Neumann, and mixed boundary conditions. Exact closed-form expressions are derived for the mean square mass density, total energy density, and thermodynamic quantities including Helmholtz free energy and entropy densities. Our analysis identifies distinct regimes, namely, a low-temperature quantum regime exhibiting characteristic power-law behavior for each boundary condition, and a high-temperature classical regime where $\hbar$-independent behavior emerges as expected. A particularly interesting finding shows that while most quantities transition naturally to classical behavior, the mean square density fluctuation requires explicit consideration of the $\hbar\to 0$ limit. The entropy density vanishes at zero temperature, in agreement with the Nernst heat theorem. Numerical analysis confirms our analytical results, particularly the asymptotic temperature behaviors and the intermediate crossover region, in which quantum and classical effects compete. This regime is governed by the energy scale $k_B T \sim \hbar u / a$, where $a$ is the distance between the planes and $u$ is the sound velocity.

\end{abstract}
\pacs{11.15.-q, 11.10.Kk} \maketitle


\section{Introduction}
The study of density fluctuations in classical liquids plays a fundamental role in understanding boundary-induced phenomena in condensed matter and quantum field theories~\cite{kittel2018introduction, lifshitz2013statistical, Dzyaloshinskii:1961sfr, bordag2009advances}. These fluctuations, which can arise from thermal and quantum effects, become particularly significant in confined geometries, where boundary conditions can strongly influence observable quantities~\cite{Ford:2008ch, Ford:2009im, deFarias:2021qdg, deFarias:2022rju}. Analogous to vacuum fluctuations in quantum field theory, in the regime where the dispersion relation is approximately linear, mass density fluctuations in fluids can be effectively modeled using phonon quantization, wherein sound waves are described by a real massless scalar field obeying the Klein-Gordon equation~\cite{lifshitz2013statistical, Ford:2009im}. This analogy holds provided the interatomic distance remains smaller than the distance scale to the boundary. It is also argued that, beyond its theoretical implications, such modified density fluctuations could be experimentally accessible through light scattering techniques~\cite{Ford:2008pae}.

The effects of modified quantum vacuum fluctuations in the phonon field are often compared with those arising from the electromagnetic field, which form the theoretical basis for the well-established Casimir effect~\cite{casimir1948attraction, bordag2009advances}. Although phonon vacuum fluctuations can produce an analogous effect, theoretical calculations show the resulting force is significantly weaker, typically suppressed by a factor of $\mathcal{O}(10^{-6})$ relative to its electromagnetic counterpart~\cite{Dzyaloshinskii:1961sfr, Ford:2009im, Ford:2008pae}. Nevertheless, several theoretical frameworks have proposed scenarios where detectable acoustic Casimir signatures might be observable, particularly in systems where finite-temperature effects are properly accounted for~\cite{bschorr1999force, larraza1998acoustic, larraza1998force, Jaskula:2012ab}. 

The aforementioned analogy also extends to the formal correspondence between the mean square mass density fluctuations of the liquid and the mean squared electric field in quantum field theory, as noted by \cite{Ford:2009im}. The influence of boundary conditions on density fluctuations has been extensively studied in various configurations. Previous work by~\cite{Ford:2008ch, Ford:2009im} examined systems with one and two parallel planes employing Neumann boundary conditions, while~\cite{deFarias:2021qdg} extended this analysis to include Dirichlet, Neumann, and mixed boundary conditions in the two-plane geometry. Building upon these foundations, we present an investigation of how two perfectly  reflecting parallel planes with Dirichlet, Neumann, and mixed boundary conditions modify both the mass density fluctuations and total energy density fluctuations in classical liquids at finite temperature. Using the framework of finite-temperature quantum field theory~\cite{birrell1984quantum}, we derive exact closed-form expressions for these physical observables. Our analysis employs the Hadamard two-point function as the fundamental tool, providing a unified treatment of thermal and boundary effects. We systematically explore the distinct behavior of these fluctuations in both the quantum ($k_BT \ll \hbar u/a$) and classical ($k_BT \gg \hbar u/a$) regimes, where $a$ represents the separation between the planes and $u$ the sound velocity.

Further developing our systematic investigation, we derive and analyze fundamental thermodynamic quantities such as the internal energy density, free energy density, and entropy density. Through detailed numerical studies, we identify and characterize three distinct temperature-dependent regimes of the confined liquid system. In the low-temperature limit, quantum fluctuations dominate the system's behavior, while the high-temperature regime is characterized by classical thermal fluctuations. Between these extremes lies an intermediate crossover region where both quantum and classical effects contribute significantly.

Our analysis reveals how the choice of the considered boundary conditions leaves distinct imprints on the thermodynamic properties of the confined liquid. The transition between quantum-dominated and classical-dominated behavior occurs at the characteristic energy scale determined by the ratio of the sound velocity $u$ to the boundary separation $a$, mediated by Planck's constant $\hbar$, that is, $k_BT\sim \frac{u\hbar}{a}$. This energy scale serves as the fundamental marker separating different physical regimes in the system. The present work establishes important connections between classical liquid systems under confinement and analogous phenomena in quantum field theory, particularly thermal versions of the Casimir effect. 

We should emphasize that, while it is employed an analogy between density fluctuations in confined fluids and quantum field theoretical vacuum fluctuations, we stress that this correspondence is phenomenological and rooted in the formal structure of field quantization in linearized hydrodynamic systems. In particular, the analogy is valid only in regimes where the dispersion relation remains approximately linear and quantum loop corrections, central in high-energy field theory, are absent. The goal is not to replicate the full vacuum structure of quantum field theory, but rather to model phononic fluctuations within a quantized framework that shares formal similarities with massless scalar fields in confined geometries.

The paper is organized as follows. In Section~\ref{sec2}, we review the fundamental aspects of phonon dynamics in liquids and present a unified derivation of the thermal Hadamard two-point function for Dirichlet, Neumann, and mixed boundary conditions. Section~\ref{sec3} contains our main analytical results, where we systematically examine the mean square mass density fluctuations, the total energy density, and derive thermodynamic quantities. For each case, we provide complete asymptotic analyzes in both temperature limits and support our findings with numerical computations. We conclude with Section~\ref{sec4}, where we summarize our principal results and discuss their implications. 



\section{Phonon in a Liquid and thermal Hadamard two-point function}
\label{sec2}
In this section, we outline the key elements of the physics describing phonon modes in a liquid, highlighting the regime where scalar quantum field theory can be applied. Next, we solve the Klein-Gordon equation for a massless real scalar field describing the phonon dynamics and consequently obtain the thermal Hadamard two-point function for the cases of 
Dirichlet, Neumann and mixed boundary conditions.

\subsection{Phonon in a Liquid}
\label{subsec2.1}
%
Before addressing our main problem, we should first introduce the minimal necessary background to understand the theory of quantum density fluctuations in a classical liquid. In this context, the fluid exhibits density fluctuations due to quantum vacuum oscillations, analogous to those in relativistic quantum field theory. These perturbations in the liquid's mass density, $\rho$, arise from quantized sound waves, known as phonons, as described by phonon theory \cite{lifshitz2013statistical}.

Mathematically, the density perturbation can be expressed as
\begin{equation}
\bar{\rho} = \rho - \rho_0,
\end{equation}
where \(\rho_0\) is a constant mean mass density characterizing the equilibrium value of $\rho$. Since the density variation is small (\(\bar{\rho} \ll \rho_0\)), the system follows a linear dispersion relation \cite{lifshitz2013statistical}
\begin{equation}
 \omega = u |k|,
\end{equation}
with \(u\) being the sound velocity in the liquid. Using the continuity equation, the perturbed mass density \(\bar{\rho}\) can then be related to the velocity field \(\vec{v}\), which for an irrotational fluid can be expressed as the gradient of a massless real scalar field, \( \phi \). As \(\vec{v}\) and $\bar{\rho}$ have the same order of magnitude, the continuity equation simplifies to
\begin{eqnarray}
\frac{\partial\bar{\rho}}{\partial t}\approx-\rho_0\nabla^2\phi,
\label{0.1}
\end{eqnarray}
where the real scalar field is referred to as the velocity potential. 

 In the following, the classical description of the fluid should be replaced with a quantum description, which can be accomplished by applying the usual quantization rules and expressing classical hydrodynamic quantities in terms of phonon annihilation and creation operators $\hat{a}_{\sigma}$, $\hat{a}^{\dagger}_{\sigma}$. These operators satisfy the canonical commutation relation 
\begin{equation}
[\hat{a}_{\sigma}, \hat{a}^{\dagger}_{\sigma'}] = \delta_{\sigma,\sigma'},
\end{equation}
and also define the phonon vacuum state $\hat{a}_{\sigma}|0\rangle=0$. Note that $\sigma$ stands for the set of quantum numbers and $\delta_{\sigma,\sigma'}$ can mean a Kronecker delta for discrete quantum numbers and a Dirac delta for continuous ones.

In the context of the quantum theory of a liquid, the density perturbation and velocity potential become operators, forming a pair of canonically conjugate variables that obey the following commutation rule:
\begin{equation}
[\hat{\bar{\rho}}(\mathbf{r}), \hat{\phi}(\mathbf{r}')] = i\hbar \delta^3(\mathbf{r} - \mathbf{r}'),
\end{equation}
given in terms of the Dirac delta function. 

Finally, the direct relation between the density perturbation and the velocity potential operators can be written as \cite{Ford:2008pae, Ford:2009im, Ford:2008ch}
\begin{equation}
\hat{\bar{\rho}}(t,\vec{r})=-\frac{\rho_0}{u^2}\dot{\hat{\phi}}(t,\vec{r}).
\label{1.4}
\end{equation}
Consequently, by substituting Eq. \eqref{1.4} into Eq. \eqref{0.1}, we are able to obtain the Klein-Gordon equation for a real massless scalar field, that is,
\begin{equation}
\left[\frac{1}{u^2}\frac{\partial^2}{\partial t^2}-\nabla^2\right]\phi=0.
\label{K_G}
\end{equation}
It is important to note that Eq.~\eqref{0.1} arises from classical hydrodynamics in the linear regime. The velocity potential $\phi$ satisfies the wave equation above due to the irrotational condition and continuity, which, in turn, justifies modeling the system as a real scalar field in an effective Minkowski background, with the sound velocity $u$ replacing the light velocity $c$, as we can see from Eq.~\eqref{K_G} \citep{lifshitz2013statistical}.

To establish a rigorous quantization scheme, we introduce a Lagrangian formulation for the effective scalar field $\phi$. Hence, the effective Lagrangian density for the fluid's phonon modes in the linear regime is
\begin{equation}
\mathcal{L} = \frac{\rho_0}{2} \left[ \frac{1}{u^2} \left( \partial_t \phi \right)^2 - \left( \nabla \phi \right)^2 \right].\label{LF}
\end{equation}
Varying the associated action, this Lagrangian leads directly to the Klein-Gordon-type wave equation in Eq.~\eqref{K_G}.

In the next subsection, we calculate the thermal Hadamard two-point function in a (3+1)-dimensional flat spacetime, considering Dirichlet, Neumann and mixed boundary conditions. We derive closed-form, analytical expressions for the thermal two-point function in each case.

\subsection{Thermal Hadamard two-point function}
\label{subsec2.2}

Having established the quantization procedure for a classical fluid in terms of phonon creation $\hat{a}_{\sigma}$ and annihilation $\hat{a}^{\dagger}_{\sigma}$ operators, thereby developing a quantum theory of liquids, we now introduce finite-temperature effects to examine their impact on fluctuation dynamics. Our analysis employs the thermal Hadamard function approach, chosen for its natural incorporation of temperature dependence in the two-point correlation function. This methodology allows systematic investigation of how thermal excitations modify the quantum vacuum fluctuations in the fluid medium.

The thermal Hadamard two-point function can be calculated by employing \cite{birrell1984quantum}
\begin{equation}
G(w, w^{\prime})=\text{Tr}[\hat{\varrho}(\phi^{*}(w^{\prime})\phi(w)+\phi(w)\phi^{*}(w^{\prime}))],
\label{2.0}
\end{equation}
where $w\equiv(t,x,y,z)$ represents spacetime coordinates and $\hat{\varrho}$ is the density matrix given by
\begin{equation}
\hat{\varrho}=Z^{-1}e^{-\beta \hat{H}},
\label{2.1}
\end{equation}
where $\beta = (k_B T)^{-1}$, with $k_B$ denoting the Boltzmann constant and $T$ the absolute temperature, while $\hat{H}$ represents the Hamiltonian operator of the system. The partition function $Z$ is written as
\begin{equation}
Z=\text{Tr}[e^{-\beta \hat{H}}].
\label{2.2}
\end{equation}

For our phonon field quantization, we employ the thermal expectation value compatible with Bose-Einstein statistics, i.e.,
\begin{equation}
\text{Tr}[\hat{\varrho}\hat{a}^+_{\sigma}\hat{a}_{\sigma^{\prime}}]=\frac{\delta_{\sigma\sigma^{\prime}}}{e^{\beta\hslash\omega_{\sigma}}-1}.
\label{2.3}
\end{equation}

%
Equipped with this formalism, we now introduce finite-temperature effects by quantizing the phonon field under three distinct boundary conditions, namely, Dirichlet, Neumann, and mixed boundary conditions, in $(3+1)$-dimensional Minkowski spacetime. For each case, we derive the exact thermal two-point function, which serves as the fundamental building block for computing finite-temperature observables. This enables systematic analysis of both the low-temperature quantum regime and high-temperature classical limit. 



\begin{figure}[h]
\includegraphics[scale=0.3]{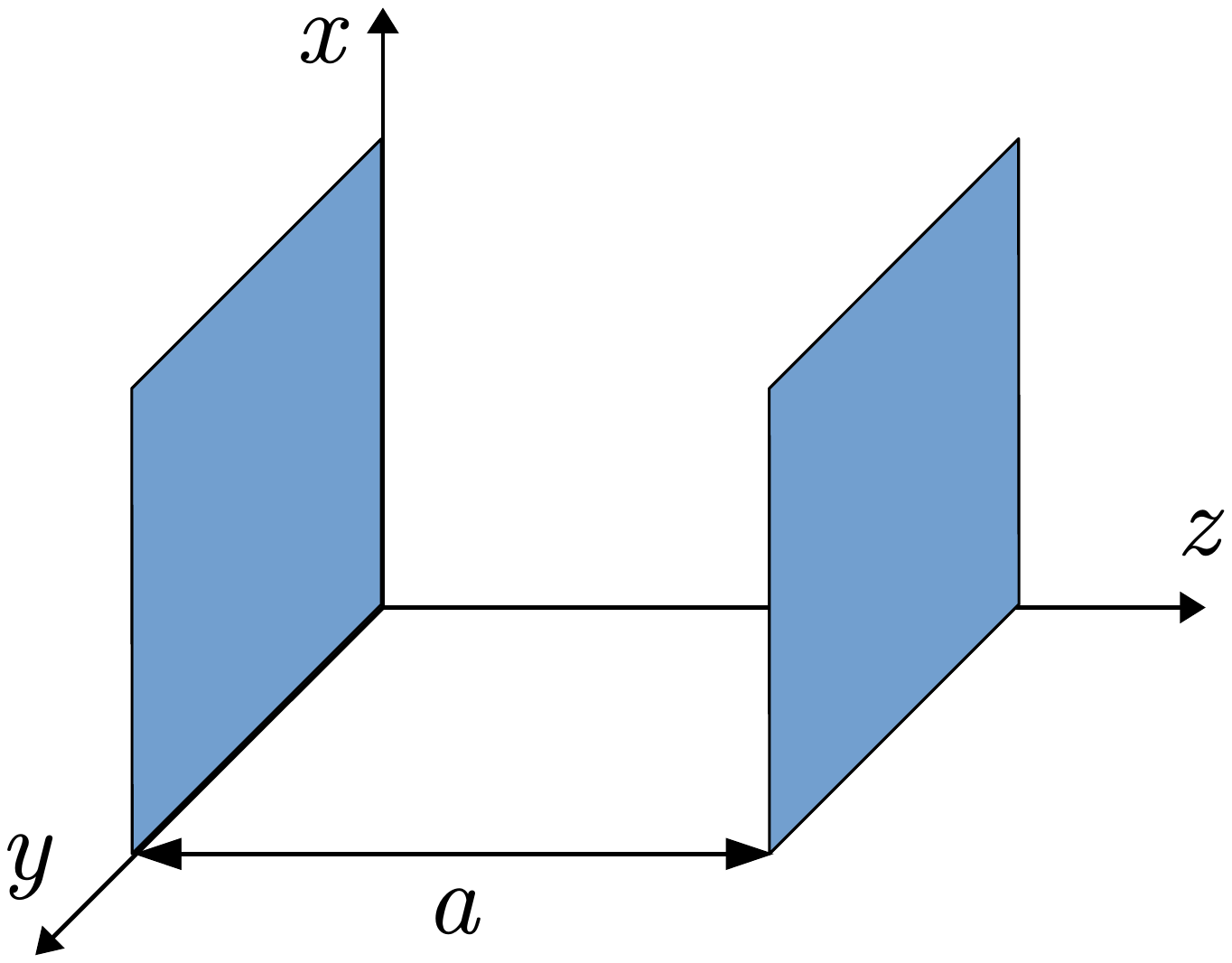}
\caption{Illustrative view of two perfectly reflecting parallel planes located at $z=0$ and $z=a$.}
\label{fig1}
\end{figure}

Now, using standard cartesian coordinates, we consider a system composed of two perfectly reflecting parallel planes located at $z=0$ and $z=a$ (see Fig.~\ref{fig1}). The real scalar field describing the phonon modes satisfies boundary conditions on these planes that modify its quantum vacuum fluctuations, leading to nonzero values for physical observables such as the fluid's mass density, as demonstrated in \cite{Ford:2008pae, Ford:2009im, Ford:2008ch,deFarias:2021qdg}. 

Our primary focus is on temperature-dependent corrections. To obtain these, we employ the mathematical framework established in the previous subsection to compute the thermal two-point function for each boundary condition, as summarized in Table~\ref{t1}.

%
\begin{table}[tbp]
\centering
\begin{tabular}{|>{\centering\arraybackslash}m{2cm}|>{\centering\arraybackslash}m{3cm}|>{\centering\arraybackslash}m{3cm}|>{\centering\arraybackslash}m{2.5cm}|>{\centering\arraybackslash}m{2.5cm}|}
\hline\hline
Type & Boundary condition & $c_{n}^2$ & $k_n$ & Solution $\varphi_n(z)$ \\ 
\hline
Dirichlet & 
  \begin{minipage}{3cm}
  \centering
  $\varphi(z=0)=0$ \\[1ex]
  $\varphi(z=a)=0$
  \end{minipage} &
  \begin{minipage}{3cm}
  \centering
  $\dfrac{\hbar u^2}{4a\pi^2\omega_{\sigma}\rho_0}$
  \end{minipage} &
  \begin{minipage}{2.5cm}
  \centering
  \vspace{0.5ex}$k_n=\dfrac{n\pi}{a}$ \\[2ex]
  $n= 1, 2, 3, \dots$
  \end{minipage} &
  \begin{minipage}{2.5cm}
  \centering
  $\sin(k_nz)$
  \end{minipage} \\
\hline
Neumann & 
  \begin{minipage}{3cm}
  \centering
  $\partial_z\varphi(z=0)=0$ \\[1ex]
  $\partial_z\varphi(z=a)=0$
  \end{minipage} &
  \begin{minipage}{3cm}
  \centering
   \vspace{0.5ex}$\dfrac{\hbar u^2}{8a\pi^2\omega_{\sigma}\rho_0}$ for $n=0$ \\[1ex]
  $\dfrac{\hbar u^2}{4a\pi^2\omega_{\sigma}\rho_0}$ for $n\geq 1$
  \end{minipage} &
  \begin{minipage}{2.5cm}
  \centering
  \vspace{0.5ex}$k_n=\dfrac{n\pi}{a}$ \\[2ex]
  $n= 0, 1, 2, \dots$
  \end{minipage} &
  \begin{minipage}{2.5cm}
  \centering
  $\cos(k_nz)$
  \end{minipage} \\
\hline
Mixed & 
  \begin{minipage}{3cm}
  \centering
   \vspace{0.5ex}$\varphi(z=0)=0$, $\partial_z \varphi(z=a)=0$ \\[1ex]
  or \\[1ex]
  $\varphi(z=a)=0$, $\partial_z \varphi(z=0)=0$
  \end{minipage} &
  \begin{minipage}{3cm}
  \centering
  $\dfrac{\hbar u^2}{4a\pi^2\omega_{\sigma}\rho_0}$
  \end{minipage} &
  \begin{minipage}{2.5cm}
  \centering
  \vspace{0.5ex}$k_n=\dfrac{(2n+1)\pi}{2a}$ \\[2ex]
  $n= 0, 1, 2, \dots$
  \end{minipage} &
  \begin{minipage}{2.5cm}
  \centering
  $\sin(k_nz)$ \\[1ex]
  or \\[1ex]
  $\cos(k_nz)$
  \end{minipage} \\
\hline\hline
\end{tabular}
\caption{Dirichlet, Neumann and mixed boundary conditions, with the their corresponding normalization constant $c_{n}^2$, discretized momentum $k_n$, and solution $\varphi_n(z)$.}
\label{t1}
\end{table}

In general, the normalized solution of the scalar field operator for the Klein-Gordon equation under the considered conditions can be expressed in terms of phonon annihilation and creation operators, $\hat{a}_{\sigma}$ and $\hat{a}_{\sigma}^{\dagger}$, as \cite{saharian2007generalized, deFarias:2021qdg}
\begin{equation}
\hat{\phi}(w)=\sum_{{\sigma}} c_{\sigma}\left(a_{\sigma}e^{-i\omega_{\sigma} t+ik_xx+ik_yy}+a^{\dagger}_{\sigma}e^{i\omega_{\sigma} t-ik_xx-k_yy}\right)\varphi_n(z),
\label{3.1.2}
\end{equation}
where $\sigma$ now represents the set of quantum numbers $(k_x, k_y, n)$. Note that in our case, the summation symbol above is defined as
\begin{equation}\label{sumsym}
\sum_{{\sigma}}= \sum_{n}\int_{-\infty}^{\infty} d k_x \int_{-\infty}^{\infty} d k_y,
\end{equation}
with $\omega_{\sigma}^2 = u^2(k_x^2 + k_y^2 + k_n^2)$ defining the dispersion relation, and the summation range over $n$ depends on the boundary condition type, as specified in Table~\ref{t1}.

We emphasize that the quantization here follows the standard procedure for linearized excitations in a medium (phonons), and does not imply the existence of full quantum vacuum loop contributions as in relativistic quantum field theory. The absence of interactions implies that our field does not generate internal loops or non-trivial vacuum diagrams. As such, our quantized scalar field exhibits thermal and boundary-induced mode modifications, but does not mimic the richer structure of quantum field theory vacua involving closed loop Feynman diagrams or virtual particle exchanges.

By substituting the field operator \eqref{3.1.2} into Eq.~\eqref{2.0} and using the result from \eqref{2.3}, we can write the thermal Hadamard two-point function in the form
\begin{eqnarray}
G(w, w^{\prime})=\sum_{{\sigma}}c_{\sigma}e^{i\vec{k}\cdot\Delta \vec \rho}
\left[\left(e^{-i\omega_{\sigma}\Delta t}+e^{i\omega_{\sigma}\Delta t}\right)+2\left(\frac{e^{-i\omega_{\sigma}\Delta t}+e^{i\omega_{\sigma}\Delta t}}{e^{\beta\hslash\omega_{\sigma}}-1}\right)\right]\varphi_n(z)\varphi_n(z'),
\label{3.1.3}
\end{eqnarray}
where $\Delta \vec \rho = \vec \rho - \vec \rho'= (\Delta x, \Delta y)$, with $\Delta x=x-x'$, $\Delta y=y-y'$, $\Delta t=t-t'$ and $\vec{k}=(k_x,k_y)$.

The two-point function in Eq.~\eqref{3.1.3} consists of two distinct components, allowing us to express it as
\begin{equation}
G(w, w^{\prime}) = G_0(x, x^{\prime}) + G_T(x, x^{\prime}),
\label{3.1.4}
\end{equation}
where the first term on the r.h.s. represents the zero-temperature Hadamard two-point function, while the second term accounts for thermal effects. Our focus is on the latter term, as the non-thermal contribution has already been thoroughly investigated in Ref.~\cite{deFarias:2021qdg}. From Eq.~\eqref{3.1.3}, the thermal component is given by
\begin{align}
G_T(w, w^{\prime})=2\sum_{{\sigma}}c_{\sigma}\frac{(e^{-i\omega_{\sigma}\Delta t}+e^{i\omega_{\sigma}\Delta t})}{e^{\beta\hslash\omega_{\sigma}}-1}e^{i\vec{k}\cdot\Delta \vec x}\varphi_n(z)\varphi_n(z').
\label{3.1.5}
\end{align}

Let us now employ polar coordinates to evaluate the integrals over momenta $k_x$ and $k_y$ (see Eq.~\eqref{sumsym}). This coordinate transformation yields $dk_xd k_y = kdkd\theta$ and $\vec{k}\cdot\Delta\vec{\rho} = k\Delta\rho\cos\theta$, where $\Delta\rho = \sqrt{\Delta x^2 + \Delta y^2}$ is the magnitude of the vector $\Delta\vec{\rho}$ and $k = \sqrt{k_x^2 + k_y^2}$ represents the wavevector magnitude. The angular integration produces $2\pi J_0(k\Delta\rho)$ in terms of the Bessel function of the first kind $J_0(k\Delta\rho)$. The resulting thermal two-point function can be expressed as
%
\begin{eqnarray}
G_T(w, w^{\prime})=4\pi\int_0^{\infty} dk \;k \; J_0(k\Delta\rho) \sum_{\delta=\pm 1}\sum_{j=1}^{\infty}\sum_{n}c_{\sigma} e^{-i\omega_{\sigma}\Delta T_{j,\delta}}\varphi_n(z)\varphi_n(z'),
\label{3.1.6}
\end{eqnarray}
where $\delta = \pm 1$ corresponds to positive and negative energy frequencies, $\Delta T_{j,\delta} = \delta\Delta t - i\beta\hbar j$ represents the complex time separation, and we have employed the identity
\begin{equation}
(e^{y}-1)^{-1}=\sum_{j=1}^{\infty}e^{-jy}.
\end{equation}

The same mathematical operations involving the $k$-integral and $n$-summation appearing in Eq.~\eqref{3.1.6} have been thoroughly analyzed in Ref.~\cite{JBFerreira:2023zwg} using the Abel-Plana formula \cite{saharian2007generalized} for Dirichlet, Neumann and mixed boundary conditions. Therefore, we can express the final result in compact form as
\begin{eqnarray}
G_T(w,w^{\prime})=\frac{\hbar u}{2\pi^2\rho_0} \sum_{\delta=\pm 1}\sum_{j=1}^{\infty} \sum_{\ell=-\infty}^{\infty}\left\{\nu_{\ell}^{(\text{i})}f_{\lambda}(\Delta w) + \epsilon_{\ell}^{(\text{i})}f_{\lambda}(\Delta\bar{w}) \right\},
\label{Gtem}
\end{eqnarray}
where $\Delta w=(\Delta x, \Delta y, \Delta z)$ and $\Delta\bar{w}=(\Delta x, \Delta y, \Delta\bar{z})$, with $\Delta z = z-z'$ and $\Delta\bar{z}=z+z'$. Moreover, we also defined 
\begin{eqnarray}
f_{\lambda}(\Delta w) = \frac{1}{(\Delta z-2a\ell)^2+ \Delta x^2 + \Delta y^2 - u^2\Delta T_{j, \delta}^2},\nonumber\\
f_{\lambda}(\Delta\bar{w}) = \frac{1}{(\Delta\bar{z}-2a\ell)^2+ \Delta x^2 + \Delta y^2 - u^2\Delta T_{j, \delta}^2},
\label{funs}
\end{eqnarray}
with $\lambda=(\ell,j,\delta)$ and 
\begin{eqnarray}
\nu_{\ell}^{\text{(i)}}&=&\left[\nu_{\ell}^{\text{(D)}},\nu_{\ell}^{\text{(N)}},\nu_{\ell}^{\text{(DN)}},\nu_{\ell}^{\text{(ND)}}\right]=[+1,+1,(-1)^{\ell},(-1)^{\ell}],\nonumber \\
\epsilon_{\ell}^{\text{(i)}}&=&\left[\epsilon_{\ell}^{\text{(D)}},\epsilon_{\ell}^{\text{(N)}},\epsilon_{\ell}^{\text{(DN)}},\epsilon_{\ell}^{\text{(ND)}}\right]=[-1,+1,(-1)^{\ell+1},(-1)^{\ell}].
\label{indices}
\end{eqnarray}
Note that the coefficients above satisfy the symmetry properties $\nu_{\ell}^{\text{(i)}} = \nu_{-\ell}^{\text{(i)}}$ and $\epsilon_{\ell}^{\text{(i)}} = \epsilon_{-\ell}^{\text{(i)}}$. Furthermore, the $\ell=0$ term in the first part of the r.h.s. represents the Minkowski thermal contribution to the two-point function, which remains finite in the coincidence limit $w'\rightarrow w$, while the $\ell=0$ term in the second part corresponds to the single-plane contribution.

\section{Mean square mass density, total energy density and thermodynamic quantities}
\label{sec3}
We now proceed to compute key physical observables, beginning with the mean square mass density fluctuations of the liquid and subsequently its total energy density and thermodynamic quantities. The present section focuses specifically on these calculations.
\subsection{Mean square mass density}
\label{ssubec3.1}
%
We begin our analysis of physical observables with the mean square mass density fluctuation. As demonstrated in Refs. \cite{Ford:2008pae, Ford:2009im, Ford:2008ch, deFarias:2022rju, deFarias:2021qdg}, this quantity serves as a fundamental observable 
characterizing the liquid and, thus, has a high physical significance.

Due to the linear dependence on creation and annihilation operators, the expectation value of the liquid's mass density vanishes, $\langle \hat{\rho} \rangle = 0$, and therefore cannot characterize the system's fluctuations. The relevant physical information is instead contained in the mass density fluctuations, quantified through its dispersion. In this sense, the finite-temperature mean square mass density is accordingly given by the thermal dispersion of Eq.~\eqref{1.4} as
\begin{eqnarray}
\langle\hat{\rho}(w)\hat{\rho}(w')\rangle_T=\frac{\rho_0^2}{2u^4}\frac{\partial^2}{\partial t\partial t^{\prime}}G_T(w,w^{\prime}),
\label{4.1}
\end{eqnarray}
where $G_T(w,w^{\prime})$ is given by Eq. \eqref{Gtem} and the factor of $1/2$ arises because the two-point Wightman function $W(w,w')$ satisfies $W(w,w') = \frac{1}{2}G(w,w')$, with $G(w,w')$ being the complete Hadamard two-point function given by Eq. \eqref{3.1.3}. This relation reflects the symmetric decomposition of the two-point Hadamard function into positive and negative frequency components. In Eq.~\eqref{4.1}, we employ the vacuum expectation value notation $\langle(...)\rangle=\langle0|(...)|0\rangle $. Note also that the expression above explicitly excludes the zero-temperature contribution to the mean square mass density, as this contribution has already been analyzed in Ref.~\cite{deFarias:2021qdg} for the boundary conditions under consideration. Hence, we will focus only on the finite-temperature contribution.

Before proceeding, we note that while the form of Eq.~\eqref{4.1} resembles expressions in quantum field theory, the underlying phononic field describes collective excitations in a classical fluid medium. Consequently, the interpretation and domain of applicability of these results remain within condensed matter analog systems and not fundamental particle physics. Again, our formalism excludes loop corrections or vacuum polarization effects that would arise in interacting quantum field theories.

The closed-form expression for the thermal component of the mean square mass density is obtained by taking the coincidence limit $w'\to w$ in Eq.~\eqref{4.1} after performing the derivative operations. This yields
\begin{align}
\langle\rho^2\rangle_{\rm T}=\frac{\rho_0}{\pi^2\hslash^3\beta^4u^5} \sum_{j=1}^{\infty} \sum_{\ell=-\infty}^{\infty} \left[\nu_{\ell}^{(\text{i})}\chi_{j\ell}(0) + \epsilon_{\ell}^{(\text{i})}\chi_{j\ell}(r)\right],
\label{4.2}
\end{align}
where 
\begin{equation}
\chi_{j\ell}(r)=\frac{3j^2-(r-\ell)^2\gamma_a^2}{[j^2 + (r-\ell)^2\gamma_a^2]^3},
\label{4.3.0}
\end{equation}
with $r=\frac{z}{a}$ and 
\begin{equation}
\gamma_a=\frac{2ak_BT}{\hslash u},
\label{gamma_a}
\end{equation}
which serves as a dimensionless scaling parameter that characterizes the system's thermal behavior. For the physical observables under study, the low-temperature regime ($\gamma_a \ll 1$) is dominated by quantum vacuum fluctuations, while the high-temperature regime ($\gamma_a \gg 1$) is governed by thermal effects, characteristic of the classical limit. A crossover region emerges at the energy scale $k_BT \sim \hbar u/a$, where quantum and classical effects compete.

For the first term on the r.h.s.\ of Eq.~\eqref{4.2}, $\ell=0$ yields the Minkowski thermal contribution, which is independent of the boundary conditions, as expected, and agrees with the results in \cite{lifshitz2013statistical,deFarias:2021qdg}. This contribution is given by

\begin{equation}
\langle\rho^2\rangle^{\rm bb}_{\rm T} = \frac{\pi^2\rho_0(k_BT)^4}{30\hbar^3 u^5},
\label{MBBody}
\end{equation}
where $\nu_{0}^{(\mathrm{i})} = 1$ for all boundary condition cases listed in Table~\ref{t1}. This expression corresponds to the mean square mass density of blackbody radiation. Note that it does not admit a classical limit through formal manipulations such as taking the limit, $\hbar \rightarrow 0$, a reasonable hypothesis. In the context of Casimir effect systems, this contribution is typically subtracted via a finite renormalization procedure to isolate the correct classical limit, a dominant contribution that must be independent of $\hbar$. This prescription is fundamentally motivated by consistency with established results for material bodies within Lifshitz theory \cite{Geyer:2008wb, Lifshitz:1956zz, Bezerra:2011zz, Bezerra:2011nc}.

For the second term on the r.h.s.\ of Eq.\ \eqref{4.2}, the contribution from $\ell = 0$ corresponds to the single-plane contribution, given by
\begin{equation}
\langle\rho^2\rangle^{\rm 1p}_{\rm T} = \epsilon_{0}^{(\mathrm{i})} \frac{\hbar\rho_0}{32\pi^2 u z^4} \left[1 - (\pi\gamma_z)^3 \frac{\cosh(\pi\gamma_z)}{\sinh^3(\pi\gamma_z)}\right],
\label{1plate}
\end{equation}
where $\epsilon_{0}^{\mathrm{(i)}}$ has been defined in Eq.~\eqref{indices} and $\gamma_z=\frac{2zk_BT}{\hslash u}$, which in this case characterizes the system’s thermal behavior. In the low-temperature regime, $\gamma_z \ll 1$, a series expansion of the second term on the r.h.s. of Eq.~\eqref{1plate} reveals that the leading-order behavior matches the thermal term in Eq.~\eqref{MBBody}, but with a sign determined by the boundary condition, that is, negative for Dirichlet ($\epsilon_{0}^{\mathrm{(i)}} = -1$) and positive for Neumann ($\epsilon_{0}^{\mathrm{(i)}} = +1$). The next-to-leading-order correction scales as $\mathcal{O}\!\left((k_B T)^6\right)$. The low-temperature asymptotic behavior consistently demonstrates that the single-plane contribution to the mean square mass density vanishes as $T \to 0$. Note also that mixed boundary conditions (DN or ND) are not applicable in the single-plane case.

On the other hand, the high-temperature regime ($\gamma_z\gg 1$) of Eq.~\eqref{1plate} is dominated by the leading-order contribution
\begin{equation}
\langle\rho^2\rangle^{\rm 1p}_{\rm T} \simeq \epsilon_{0}^{(\mathrm{i})}\frac{\hbar\rho_0}{32\pi^2uz^4} + \rm{E.S.}\,,
\label{highT_1plate}
\end{equation}
where E.S. denotes exponentially suppressed terms. The expression above shows a surprising dependence on $\hbar$ but no dependence on $k_BT$, behavior contrary to what one expects in a classical regime. However, the classical limit is properly recovered only when $\hbar\rightarrow 0$, which gives $\langle\rho^2\rangle^{\rm 1p}_{\rm T} = 0$ in the high-temperature regime. For massive scalar fields, Ref.~\cite{DeLorenci:2014maa} shows that no classical contribution appears at high temperatures, in such cases, the classical limit also emerges exclusively through the $\hbar\rightarrow 0$ limit (see Eq.~(3.9) in Ref.~\cite{DeLorenci:2014maa}). 
\begin{figure}[h] 
    \centering 
    \includegraphics[width=0.45\textwidth]{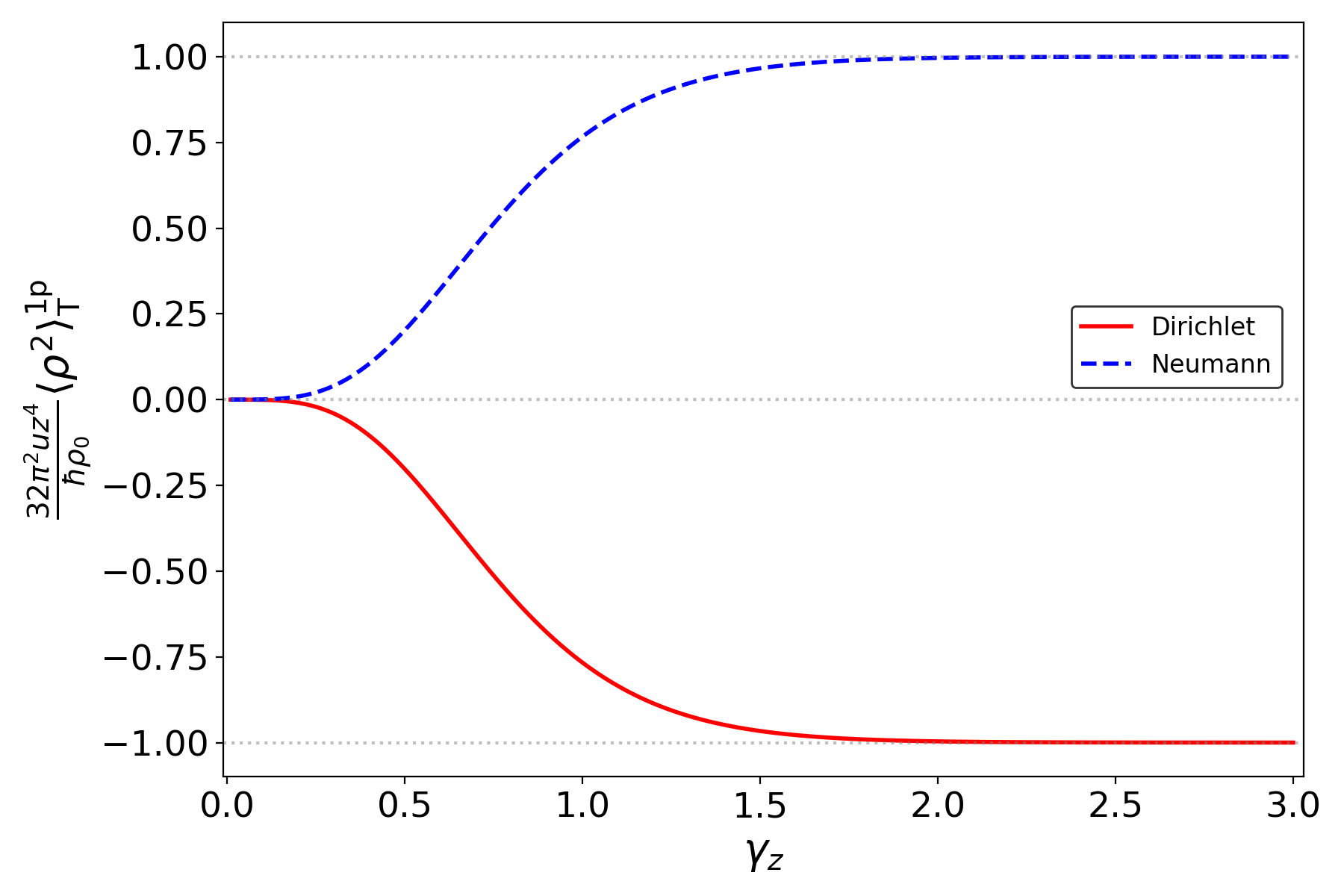}
    \includegraphics[width=0.45\textwidth]{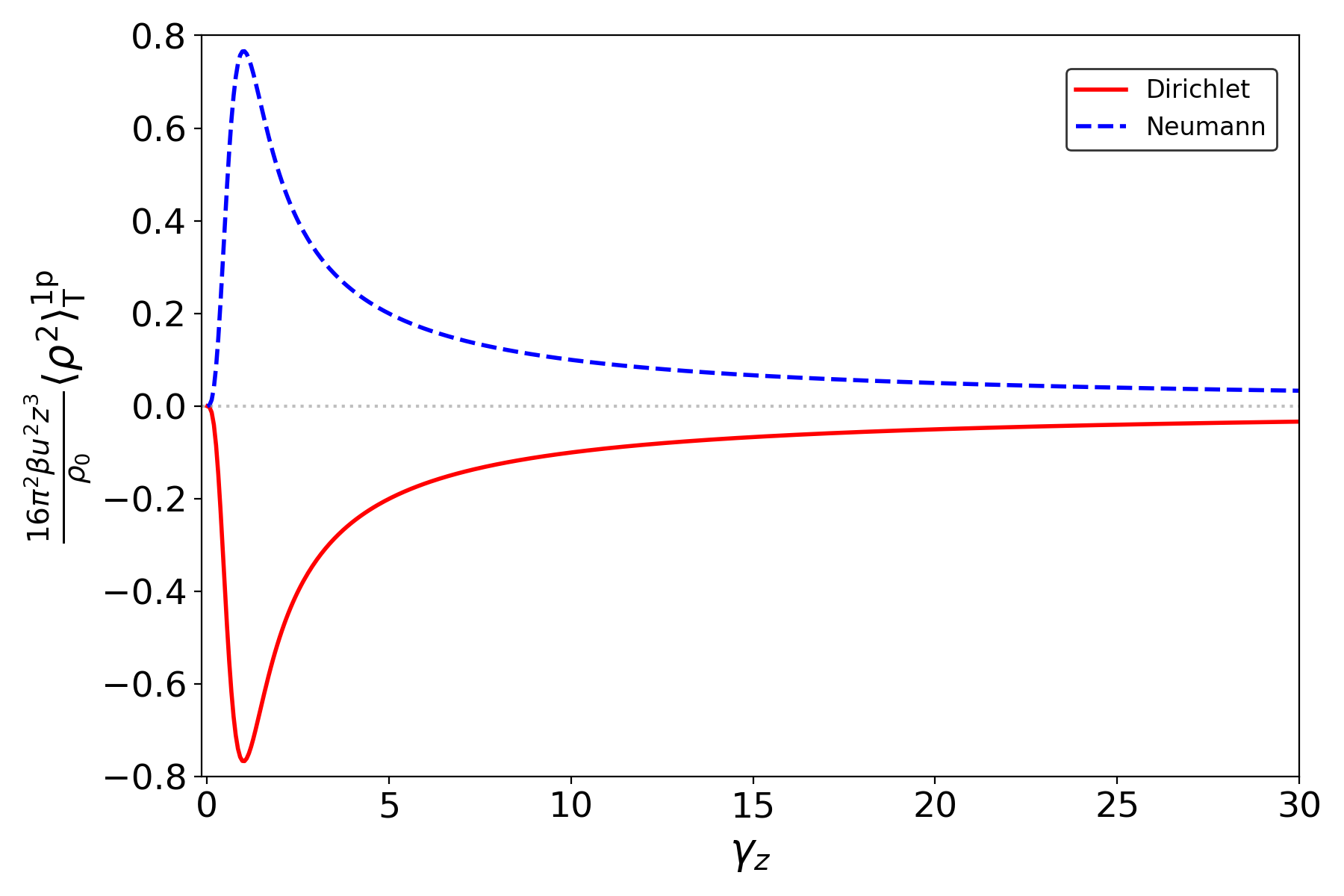} 
    \caption{The mean square mass density from Eq.~\eqref{1plate} is shown as a function of the dimensionless parameter $\gamma_z$. On the left, the plot reveals unexpected asymptotic behavior at high temperatures, while on the right, the classical limit is recovered as $\hbar \to 0$.} 
    \label{fig2} 
\end{figure}

In Fig.~\ref{fig2}, we numerically plot the curves for the expression in Eq.~\eqref{1plate} as a function of $\gamma_z$. The left plot illustrates the asymptotic behavior discussed above, where the mean square mass density approaches a constant value at high temperatures, determined by $\hbar$, as indicated by Eq.~\eqref{highT_1plate}. On the other hand, the plot on the right demonstrates how Eq.~\eqref{1plate} vanishes at high temperatures when taking $\hbar \to 0$. The resulting curves are now consistent with the expected behavior for both the low- and high-temperature regimes.

Now, by considering Eq.~\eqref{4.2} while excluding both the blackbody term~\eqref{MBBody} and the single-plane component~\eqref{1plate}, we obtain the purely two-plane contribution
\begin{align}
\langle\rho^2\rangle^{\rm 2p}_{\rm T} = \frac{\rho_0\hbar\gamma_a^4}{16\pi^2ua^4} \sum_{j=1}^{\infty} \sum_{\ell=1}^{\infty} \left\{2\nu_{\ell}^{(\mathrm{i})}\chi_{j\ell}(0) + \epsilon_{\ell}^{(\mathrm{i})}\left[\chi_{j\ell}(r) + \chi_{j\ell}(-r)\right]\right\},
\label{4.5}
\end{align}
where the function $\chi_{j\ell}(r)$ has been defined in Eq.~\eqref{4.3.0}. The double sum in Eq.~\eqref{4.5} can be computed in two different ways, each making distinct temperature regimes more apparent. For instance, performing first the sum over $\ell$ followed by the sum over $j$ reveals the low-temperature regime, while performing the sums in the reverse order (first $j$ then $\ell$) makes the high-temperature regime more evident. Both approaches yield identical results for the mean square mass density in Eq.~\eqref{4.5}, with the choice of summation order serving merely as an analytical tool to isolate different thermal behaviors.  Although neither approach allows for the analytical evaluation of both sums, performing the summation over $j$ first yields a tractable expression that we present at the end of this subsection. This enables practical numerical analysis and graphical representation through specialized computational tools. 


Let us now examine the low-temperature regime, $\gamma_a \ll 1$, where we first perform the sum over $\ell$ followed by the sum over $j$. The leading-order contribution for each boundary condition is as follows
\begin{equation}
\langle\rho^2\rangle^{\rm 2p}_{\rm T} \simeq
\begin{cases} 
-\dfrac{\rho_0\hbar}{ua^4} \dfrac{\pi^4r^2}{1512}\gamma_a^6 + \mathcal{O}\left(\gamma_a^8\right) & \text{(Dirichlet)}, \\[0.3cm]
\;\;\;\dfrac{\rho_0\hbar}{ua^4} \dfrac{\zeta(3)}{8\pi}\gamma_a^3 + \mathcal{O}\left(\gamma_a^4\right) & \text{(Neumann)}, \\[0.3cm]
\;\;\;\dfrac{\rho_0\hbar}{ua^4} \dfrac{\pi^2}{640}\gamma_a^4 + \mathcal{O}\left(\gamma_a^6\right) & \text{(DN)}, \\[0.3cm]
-\dfrac{\rho_0\hbar}{ua^4} \dfrac{\pi^2}{384}\gamma_a^4 + \mathcal{O}\left(\gamma_a^6\right) & \text{(ND)}.
\end{cases}
\label{lowTLiquidD}
\end{equation}
We observe that all expressions above vanish as $(k_B T)^n$ in the low-temperature limit, with the exponents $n = 3, 4,$ and $6$ corresponding to Neumann, mixed (DN/ND), and Dirichlet boundary conditions, respectively. This temperature dependence aligns with our theoretical expectations for these boundary conditions.

In contrast, the high-temperature limit, $\gamma_a \gg 1$, requires that we first perform the sum over $j$ followed by the sum over $\ell$. In this case, the leading-order contribution is given by
\begin{align}
\langle\rho^2\rangle^{\rm 2p}_{\rm T} \simeq \frac{\rho_0\hbar}{32\pi^2ua^4} \sum_{\ell=1}^{\infty} \left\{\frac{2\nu_{\ell}^{(\text{i})}}{\ell^4}  + \epsilon_{\ell}^{(\text{i})}\left[\frac{1}{(\ell - r)^4} + \frac{1}{(\ell + r)^4} \right]\right\} + \rm{E.S.}\,,
\label{eq:high_rho}
\end{align}
which despite being the high-temperature limit, the expression retains its quantum character through the explicit $\hbar$ dependence. This again appears to contradict our conventional expectation that high temperatures should yield classical behavior. Nevertheless, the proper classical limit emerges when we formally set $\hbar \to 0$, consistent with the procedure employed previously.
\begin{figure}[h] 
    \centering 
    \includegraphics[width=0.46\textwidth]{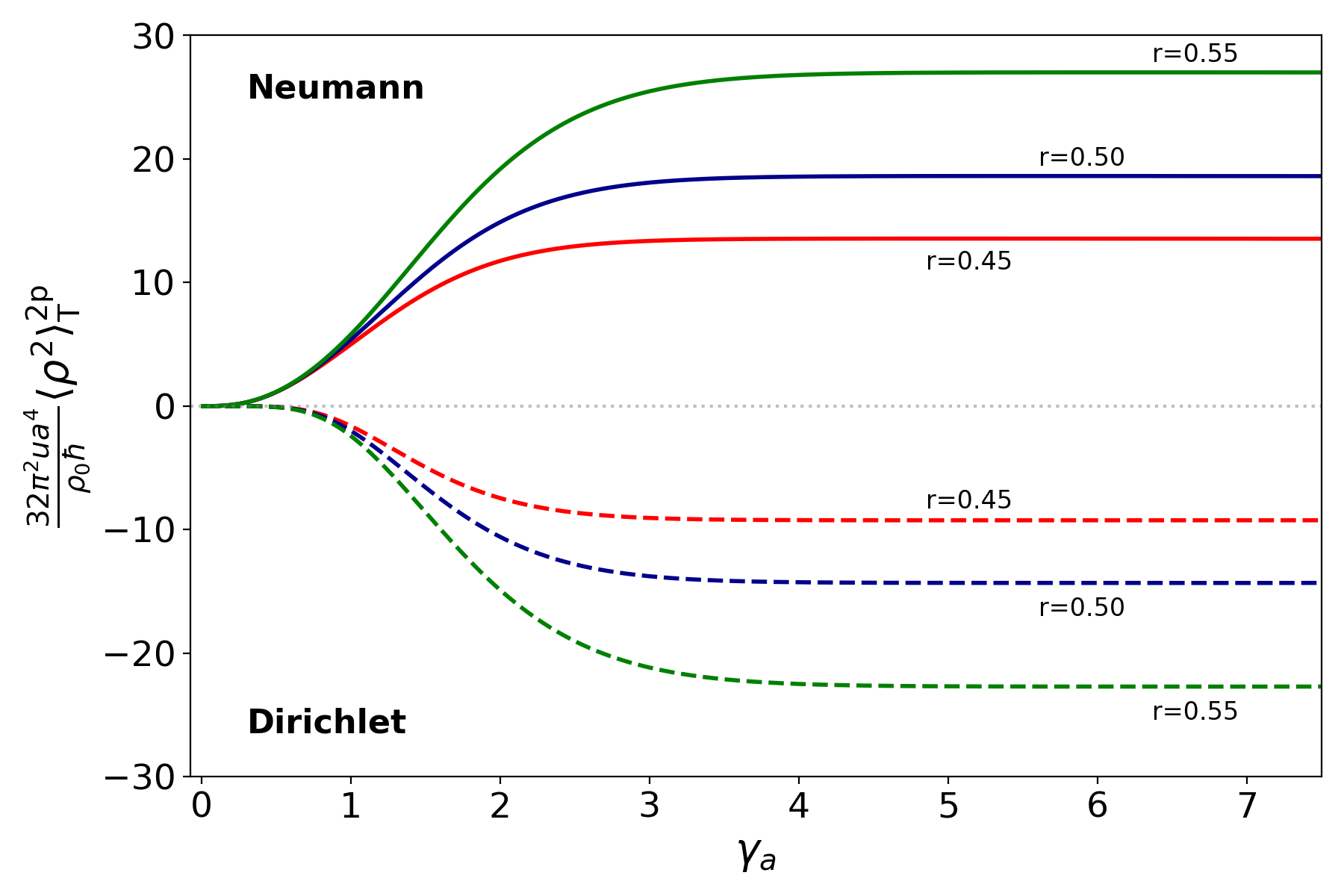}
    \includegraphics[width=0.46\textwidth]{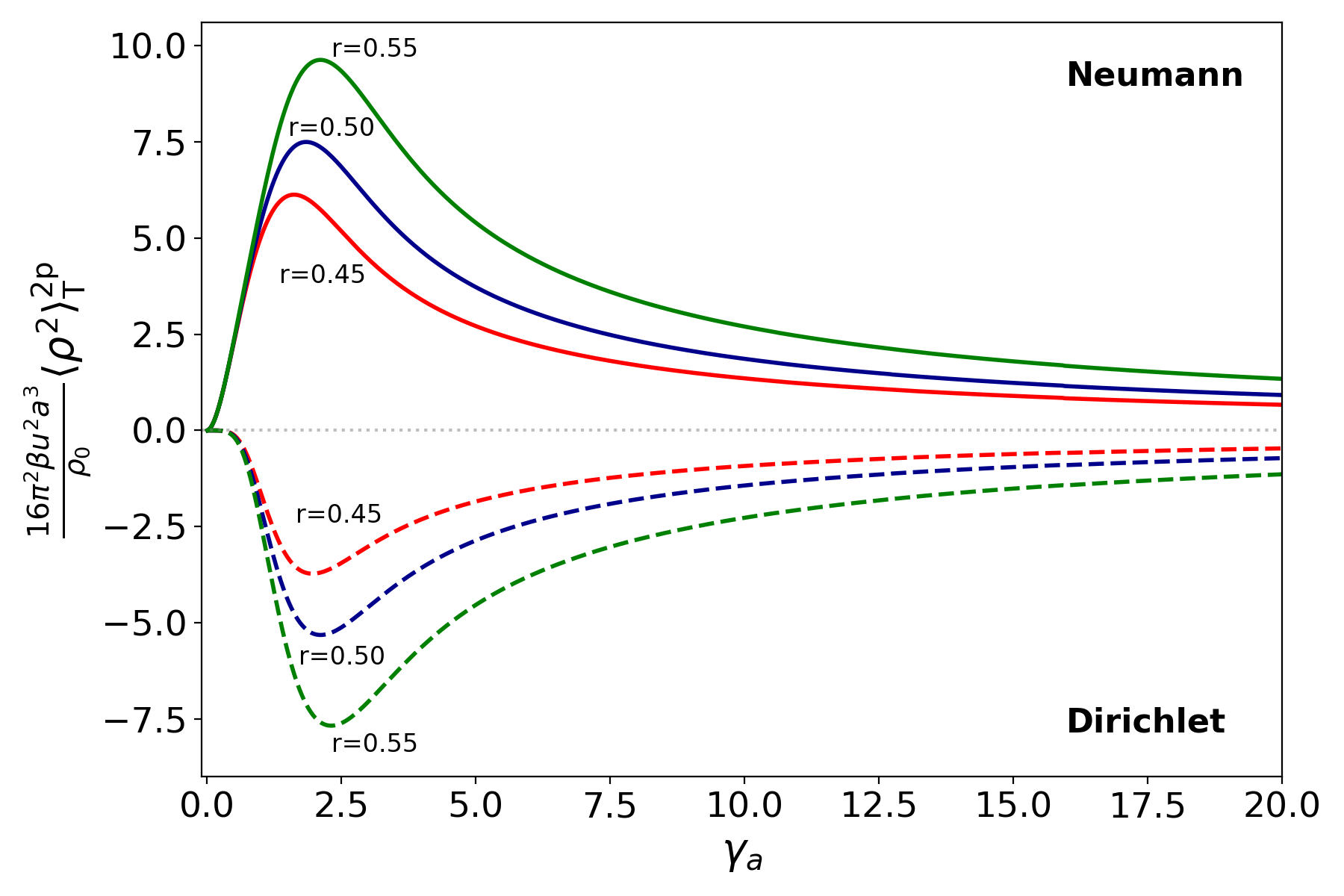} 
    \includegraphics[width=0.46\textwidth]{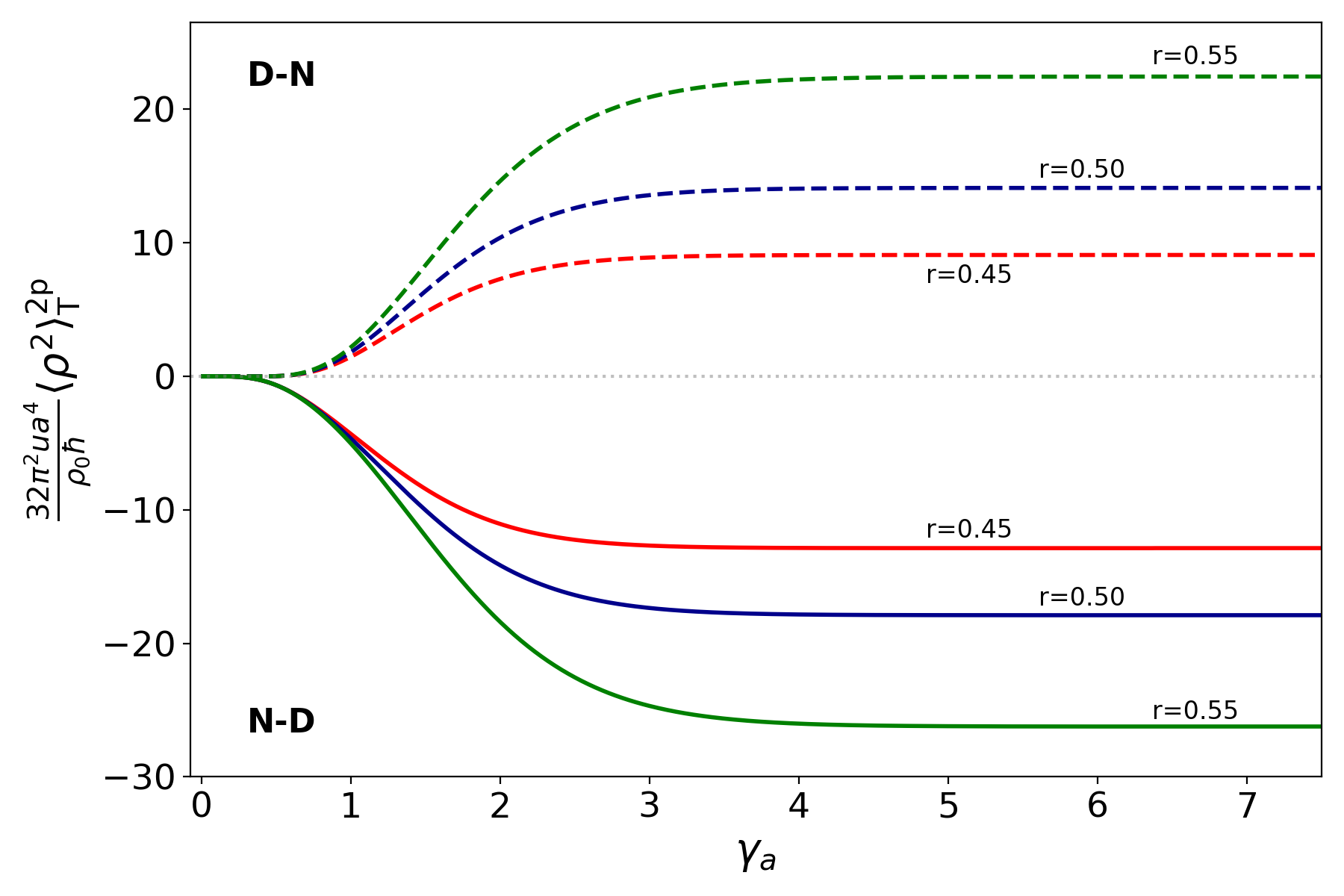}
      \includegraphics[width=0.46\textwidth]{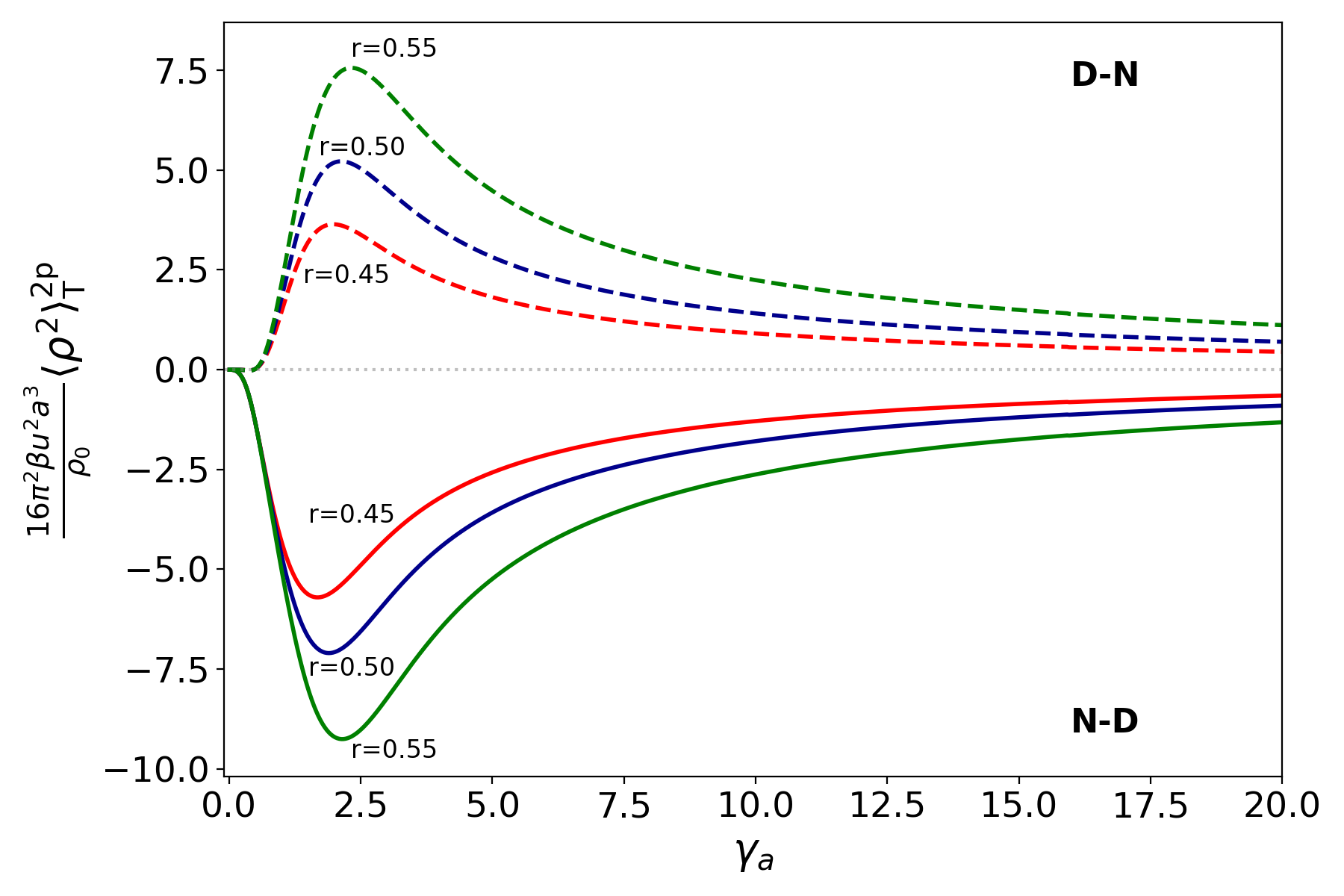}
    \caption{The mean square mass density from Eq.~\eqref{eq:sum_ell} is plotted as a function of the dimensionless parameter $\gamma_a$, for various values of $r$ and for each boundary condition case listed in Table~\ref{t1}. The left plots display unexpected asymptotic behavior in the high-temperature regime, while the right plots show the recovery of the classical limit when $\hbar \to 0$.} 
    \label{fig3} 
\end{figure}

We now derive a computationally tractable expression for the mean square mass density \eqref{4.5}, suitable for numerical evaluation. The calculation strategy involves performing the summation over $j$ before addressing the $\ell$-summation, as the reverse order would lead to more analytically complicated expressions. This approach yields 
\begin{equation}
\langle\rho^2\rangle^{\rm 2p}_{\rm T} = \frac{\rho_0\hbar}{32\pi^2ua^4}\sum_{\ell=1}^{\infty} 
\Big\{
2\nu_{\ell}^{(\text{i})}\beta_{T}(\ell) 
+ \epsilon_{\ell}^{(\text{i})}\Big[\beta_{T}(\ell + r)  + \beta_{T}(\ell - r)\Big]
\Big\},
\label{eq:sum_ell}
\end{equation}
where we have introduced the auxiliary function 
\begin{equation}
\beta_{T}(v) = \frac{1 - (\pi \gamma_a v)^3\coth(\pi \gamma_a v)\operatorname{csch}^2(\pi \gamma_a v)}{v^4}.
\label{eq:beta_func}
\end{equation}

Fig.~\ref{fig3} shows the dimensionless mean square mass density from Eq.~\eqref{eq:sum_ell} as a function of $\gamma_a$, presented for different values of $r$ and for each boundary condition case. As in the single-plane scenario, the left plot demonstrates a high-temperature limit that depends on Planck's constant, while the right plot reveals that Eq.~\eqref{eq:sum_ell} vanishes when $\hbar\to 0$, thus recovering the classical limit. Note also that the asymptotic low-temperature behavior described by Eq.~\eqref{lowTLiquidD} is consistent with the curves shown in Fig.~\ref{fig3}. That is, the mean square mass density in the two-plane scenario vanishes as $T \to 0$, as expected.

\subsection{Total energy density of the liquid}
\label{subsec3.2}
%
The energy density in the form of phonon waves in a liquid is more accurately obtained by making use of the Hamiltonian density operator that incorporates both internal and kinetic energy contributions \cite{lifshitz2013statistical}. This quantity can be elegantly expressed in terms of the massless scalar field operator $\hat{\phi}$ as

\begin{equation}
\hat{\mathcal{H}} = \frac{\rho_0}{2}\left[(\nabla\hat{\phi})^2 + \frac{1}{u^2}(\partial_t\hat{\phi})^2\right],
\label{eq:hamiltonian_density}
\end{equation}
where the first and second terms represent the kinetic and internal energy density operators, respectively \cite{lifshitz2013statistical}. Note that the Hamiltonian formulation above can be interpreted as corresponding to a field satisfying the Euler-Lagrange equations derived from the effective Lagrangian density in Eq.~\eqref{LF}.

Furthermore, the vacuum expectation value of the Hamiltonian density operator can be rewritten using the Hadamard function \eqref{3.1.4}. By considering only the finite-temperature contribution, we may employ the point-splitting method to express the vacuum expectation value of Eq.~\eqref{eq:hamiltonian_density} in the following form:
\begin{equation}
\langle \mathcal{H}\rangle_{\mathrm{T}} = \lim_{w^{\prime}\to w} \frac{\rho_0}{2 \times 2} \left[ \frac{1}{u^2} \partial_t \partial_{t^{\prime}} + \partial_x \partial_{x^{\prime}} + \partial_y \partial_{y^{\prime}} + \partial_z \partial_{z^{\prime}} \right] G_{T}(w, w^{\prime}),
\label{eq:mean_hamiltonian}
\end{equation}
which represents the total energy density of the liquid. Here, an additional factor of $1/2$ has been introduced to account for the symmetric decomposition of the Hadamard two-point function into positive- and negative-frequency components, analogous to the procedure in Eq.~\eqref{4.1}. 

By substituting Eq.~\eqref{Gtem} for the finite-temperature contribution of the Hadamard function into Eq.~\eqref{eq:mean_hamiltonian}, we obtain
\begin{align}
\langle \mathcal{H}\rangle_{\mathrm{T}} &= \frac{1}{\pi^2(\hbar u)^3\beta^4}\sum_{j=1}^{\infty}\sum_{\ell=-\infty}^{\infty}
\left\{\nu_{\ell}^{(\mathrm{i})}\chi_{j\ell}(0) + 2\epsilon_{\ell}^{(\mathrm{i})}\vartheta_{j\ell}(r) \right\},
\label{eq:energy_density}
\end{align}
where we have defined
\begin{equation}
\vartheta_{j\ell}(r) = \frac{1}{[j^2 + \gamma_a^2(r + \ell)^2]^2}.
\label{theta}
\end{equation}

The energy density in Eq.~\eqref{eq:energy_density} can be separated into three distinct contributions: the energy density of blackbody radiation, the single-plane and two-plane contributions. The first term on the right-hand side of Eq.~\eqref{eq:energy_density} (with $\ell = 0$) yields the blackbody contribution given by

\begin{eqnarray}
\langle \mathcal{H}\rangle^{\rm bb}_{\rm T} &=& \frac{u^2}{\rho_0} \langle \rho^2\rangle^{\rm bb}_{\rm T}, \nonumber\\
&=& \frac{\pi^2(k_BT)^4}{30(\hbar u)^3},
\label{BBed}
\end{eqnarray}
where we have used Eq.~\eqref{MBBody}.

Additionally, the second term in Eq.~\eqref{eq:energy_density} provides the single-plane contribution for $\ell=0$. Consequently, the sum over $j$ of the function $\vartheta_{j0}(r)$ from Eq.~\eqref{theta} may be evaluated using the analytic continuation of the Epstein-Hurwitz zeta function \cite{Elizalde:1995hck,Elizalde_hiroshima}:
\begin{align}
\zeta_{\rm EH}(s, q) &= \sum_{k=1}^{\infty} \frac{1}{[k^2+q^2]^s} \nonumber \\
&= -\frac{q^{-2s}}{2} + \frac{\pi^{\frac{1}{2}}\Gamma(s-1/2)}{2\Gamma(s)}q^{1-2s} \nonumber \\
&\quad + \frac{2\pi^sq^{\frac{1}{2}-s}}{\Gamma(s)}\sum_{n=1}^{\infty} n^{s-\frac{1}{2}}K_{s-\frac{1}{2}}(2\pi nq),
\label{E-H}
\end{align}
where $K_{\alpha}(x)$ is the modified Bessel function of the second kind (Macdonald function) and $\Gamma(s)$ is the gamma function \cite{abramowitz1970handbook,gradshtein2007}. This yields
\begin{align}
\langle \mathcal{H}\rangle^{\rm 1p}_{\rm T} = -\epsilon_{0}^{(\mathrm{i})}\frac{\hbar u}{16\pi^2z^4}\left[1 - \frac{\pi\gamma_z}{2} - \frac{\pi\gamma_z\left(1 + 2\pi\gamma_z - e^{-2\pi\gamma_z}\right)}{4\sinh^2(\pi\gamma_z)}\right],
\label{total_energy1p}
\end{align}
where, according to Eq.~\eqref{indices}, the coefficient $\epsilon_{0}^{(\mathrm{i})} = -1$ for Dirichlet and $\epsilon_{0}^{(\mathrm{i})} = +1$ for Neumann boundary conditions. Thus, the difference between the two conditions amounts to a sign change. 

For the asymptotic temperature regimes, we observe that at high temperatures, $\gamma_z=\frac{2zk_BT}{\hbar u} \gg 1$, the expression is dominated by the second term on the r.h.s., yielding
\begin{align}
\langle \mathcal{H}\rangle^{\rm 1p}_{\rm T} \simeq \epsilon_{0}^{(\mathrm{i})}\frac{k_B T}{16\pi z^3} + \mathrm{E.S.}\;.
\label{total_energy1p_highT}
\end{align}
As evident, this represents a classical term since it is independent of $\hbar$, consistent with the expectation that quantum effects become negligible at high temperatures.

In contrast, for the low-temperature regime, $\gamma_z \ll 1$, the energy density in Eq.~\eqref{total_energy1p} for a single plane is dominated by the leading quantum contribution
\begin{align}
\langle \mathcal{H}\rangle^{\rm 1p}_{\rm T} \simeq \epsilon_{0}^{(\mathrm{i})}\frac{(k_B T)^2}{12\hbar u z^4} + \mathcal{O}\left((k_B T)^3\right),
\label{total_energy1p_lowT}
\end{align}
where the explicit $\hbar$-dependence clearly demonstrates the quantum nature of this low-temperature behavior. 
\begin{figure}[h] 
\centering 
\includegraphics[width=0.46\textwidth]{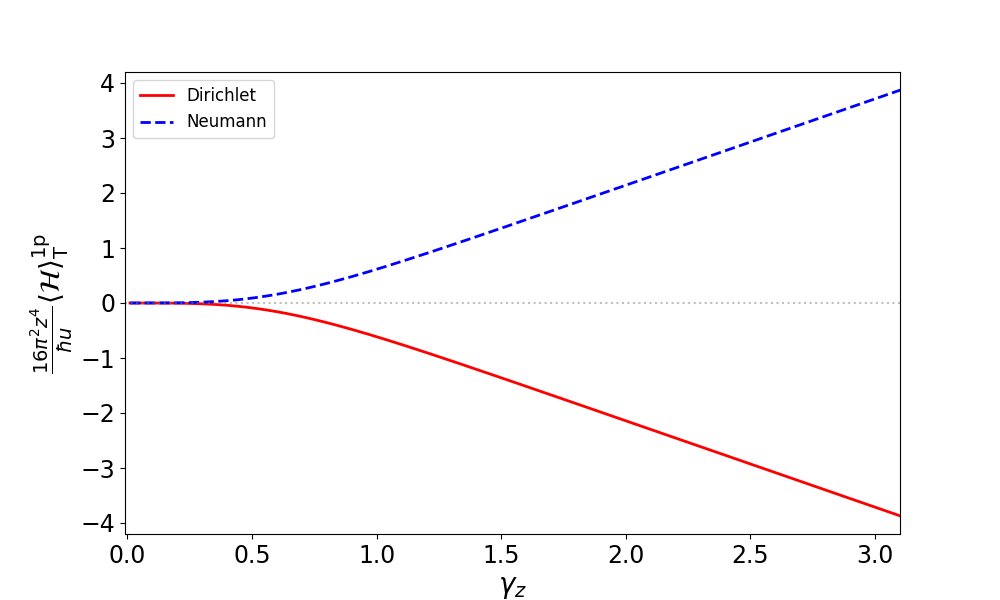}
 \caption{Total energy density from Eq.~\eqref{total_energy1p} plotted as a function of $\gamma_z$.}
\label{fig4} 
\end{figure}

Fig.~\ref{fig4} presents the dimensionless total energy density from Eq.~\eqref{total_energy1p} as a function of $\gamma_z$. The curves exhibit the asymptotic behaviors described by Eqs.~\eqref{total_energy1p_highT} and \eqref{total_energy1p_lowT}. Specifically, the energy density varies linearly with temperature $T$ in the high-temperature regime, while approaching zero in the low-temperature limit, as confirmed by Eqs.~\eqref{total_energy1p_highT} and \eqref{total_energy1p_lowT}, respectively.

Finally, the third contribution in Eq.~\eqref{eq:energy_density} corresponds to that of two planes. This is obtained by subtracting the blackbody and single-plane contributions, respectively, in Eqs. \eqref{BBed} and \eqref{total_energy1p}. This provides
\begin{align}
\langle \mathcal{H}\rangle^{\rm 2p}_{\rm T} &= \frac{\hbar u\gamma_a^4}{8\pi^2 a^4}\sum_{j=1}^{\infty}\sum_{\ell=1}^{\infty}
\left\{
\nu_{\ell}^{(\mathrm{i})}\chi_{j\ell}(0) 
+ \epsilon_{\ell}^{(\mathrm{i})}\left[\vartheta_{j\ell}(r) + \vartheta_{j\ell}(-r)\right]
\right\}.
\label{2plate}
\end{align}

To analyze the low-temperature regime, $\gamma_a \ll 1$, we first perform the summation over $\ell$ followed by the summation over $j$. This yields the leading contributions for each boundary condition case in Table~\ref{t1}, that is,

\begin{equation}
\langle \mathcal{H}\rangle^{\rm 2p}_{\rm T} \simeq 
\begin{cases} 
-\dfrac{\hbar u}{a^4} \dfrac{\pi^2}{1440}\gamma_a^4 + \mathcal{O}\left(\gamma_a^6\right) & \text{(Dirichlet)}, \\[0.5em]
\;\;\;\dfrac{\hbar u}{a^4} \dfrac{\zeta(3)}{8\pi}\gamma_a^3 + \mathcal{O}\left(\gamma_a^4\right) & \text{(Neumann)}, \\[0.5em]
\;\;\;\dfrac{\hbar u}{a^4} \dfrac{\pi^2}{1152}\gamma_a^4 + \mathcal{O}\left(\gamma_a^6\right) & \text{(DN)}, \\[0.5em]
-\dfrac{\hbar u}{a^4} \dfrac{11\pi^2}{5760}\gamma_a^4 + \mathcal{O}\left(\gamma_a^6\right) & \text{(ND)},
\end{cases}
\label{lowTEnergyD}
\end{equation}
in accordance with what is expected once again.

On the other hand, in the high-temperature limit, $\gamma_a\gg 1$, we should first perform the sum over $j$ and then the sum over $\ell$. This leads to:
\begin{eqnarray}
\langle \mathcal{H}\rangle^{\rm 2p}_{\rm T} \simeq \frac{k_BT}{16\pi a^3} \sum_{\ell=1}^{\infty} \epsilon_{\ell}^{(\text{i})} \left[ \frac{1}{(\ell + r)^3} + \frac{1}{(\ell - r)^3} \right] + \rm{E.S.}\,,
\label{highT}
\end{eqnarray}
where the sum above results in known special functions. We can see that the classical contribution remains dominant in the high-temperature regime in agreement with theoretical predictions.

Similar to the previous subsection, we can perform the sum over $j$ in Eq. \eqref{2plate} to obtain a more practical expression for computational analysis. The sum aforementioned provides
\begin{align}
\langle \mathcal{H}\rangle^{\rm 2p}_{\rm T} &= \frac{\hbar u}{32\pi^2 a^4}\sum_{\ell=1}^{\infty}
\left\{2\nu_{\ell}^{(\mathrm{i})}\beta_T(\ell) 
+ \epsilon_{\ell}^{(\mathrm{i})}\left[\Lambda_{T}(\ell + r) + \Lambda_{T}(\ell - r)\right]
\right\},
\label{2plate}
\end{align}
where the function $\beta_T(v)$ has been defined in \eqref{eq:beta_func} and
\begin{equation}
\Lambda_{T}(v) = \frac{-2 + \gamma_a \pi v \big[\coth\big(\gamma_a\pi v\big) + \gamma_a \pi v \operatorname{csch}^2\big(\gamma_a \pi v\big)\big]}{v^4}.
\label{Lambda_func}
\end{equation}

\begin{figure}[h] 
    \centering 
    \includegraphics[width=0.46\textwidth]{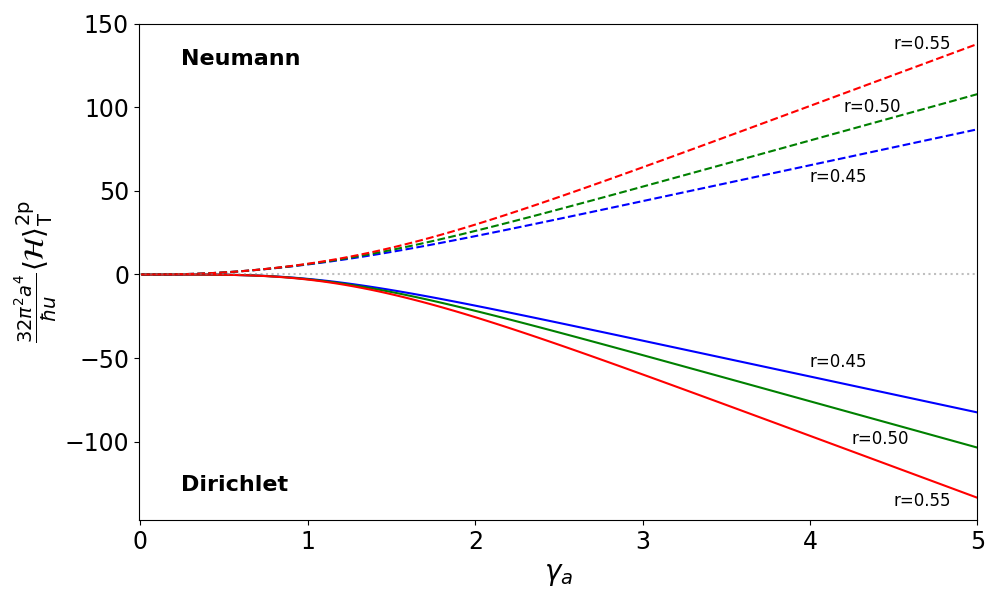}
    \includegraphics[width=0.45\textwidth]{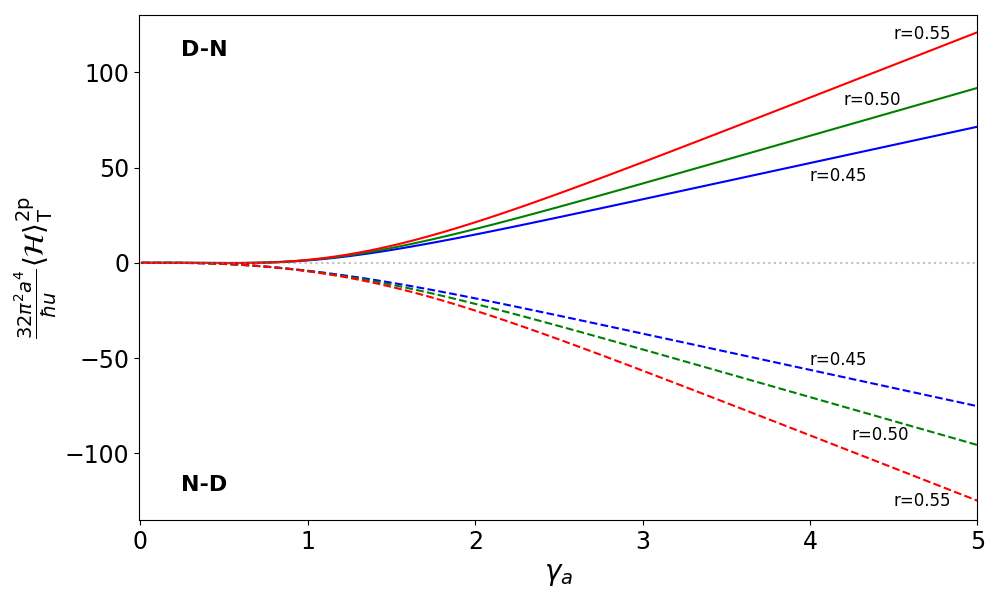}
    \caption{The total energy density from Eq.~\eqref{2plate} is plotted as a function of the dimensionless parameter $\gamma_a$, for various values of $r$ and for each boundary condition case listed in Table~\ref{t1}.}
    \label{fig5} 
\end{figure}

Fig.~\ref{fig5} displays the dimensionless total energy density from Eq.~\eqref{2plate} as a function of $\gamma_a$ for various values of $r$ and for each boundary condition considered. The left plot shows the curves for Dirichlet and Neumann boundary conditions, while the right plot presents those for DN and ND configurations. Both plots clearly demonstrate agreement with the asymptotic behaviors described by Eqs.~\eqref{lowTEnergyD} and \eqref{highT}.

\subsection{Thermodynamic quantities}

The first thermodynamic quantity we consider is the internal energy density, which has a direct relation to the mean square mass density, exhibiting identical behavior up to dimensional constants. Hence, the internal energy density at finite temperature is written as \cite{lifshitz2013statistical}
\begin{eqnarray}
\mathcal{U}_T=\frac{u^2}{\rho_0}\langle\rho^2\rangle_T,
\label{4.7}
\end{eqnarray}
which is in fact the second term on the r.h.s. of Eq. \eqref{eq:hamiltonian_density}. In the above expression, just as in the mean square mass density, there are contributions from the blackbody radiation term, Eq. \eqref{MBBody}, from the single-plane term, Eq. \eqref{1plate}, and from the two-plane term, Eq. \eqref{4.5}. The black-body radiation contribution from \eqref{4.7} agrees with established results from \cite{lifshitz2013statistical}, as expected. The zero-temperature contribution to $\langle\rho^2\rangle$, previously investigated in 
\cite{deFarias:2021qdg}, represents a pure boundary effect on the internal energy density. 
This quantum boundary term, associated with the vacuum state of the phonon field, has no 
analogue in conventional thermodynamics.

The Helmholtz free energy density can be derived from Eq.~\eqref{4.7} through the 
fundamental thermodynamic relation
\begin{equation}
\mathcal{U}_T = -T^2\frac{\partial}{\partial T}\left(\frac{\mathcal{F}_T}{T}\right).
\label{FreeE0}
\end{equation}
By integrating the above expression, we find
\begin{equation}
\mathcal{F}_T = -T\frac{u^2}{\rho_0^2}\int \frac{dT}{T^2}\langle\rho^2\rangle_T + CT,
\label{FreeE1}
\end{equation}
where the integration constant $C$ must vanish to ensure the entropy $S \rightarrow 0$ as $T \rightarrow 0$, in accordance with the third law of thermodynamics (Nernst heat theorem) \cite{Landau:1980mil}, as shown later.

Substituting Eq.~\eqref{4.2} into Eq.~\eqref{FreeE1} yields
\begin{equation}
\mathcal{F}_T = -\frac{1}{\pi^2(\hbar u)^3\beta^4}\sum_{j=1}^{\infty}\sum_{\ell=-\infty}^{\infty}\left[\nu_{\ell}^{(\text{i})}\vartheta_{j\ell}(0) + \epsilon_{\ell}^{(\text{i})} \vartheta_{j\ell}(r)\right],
\label{FreeE}
\end{equation}
where $\vartheta_{j\ell}(r)$ has been defined in Eq.~\eqref{theta}. The $\ell=0$ component of 
$\vartheta_{j\ell}(0)$ reproduces the free energy density of blackbody radiation, 
$\mathcal{F}^{\mathrm{bb}}_T = -\frac{1}{3}\langle \mathcal{H}\rangle^{\mathrm{bb}}_{\mathrm{T}}$ 
\cite{lifshitz2013statistical, Bezerra:2011zz}. Similarly, the $\ell=0$ term from 
$\vartheta_{j\ell}(r)$ corresponds to the single-plane contribution. When summed over $j$ 
using Eq.~\eqref{E-H}, this yields $\mathcal{F}^{\mathrm{1p}}_T = -\frac{1}{2}\langle 
\mathcal{H}\rangle^{\mathrm{1p}}_{\mathrm{T}}$, exhibiting the same asymptotic temperature 
dependence as $\langle\mathcal{H}\rangle^{\mathrm{1p}}_{\mathrm{T}}$ in Eq.~\eqref{total_energy1p}.

The two-plane contribution for the free energy density still remains to be discussed. It is obtained from Eq. \eqref{FreeE} by subtracting both the free energy density of blackbody and the single-plane contributions. Thus, we have
\begin{equation}
\mathcal{F}^{\rm 2p}_{\rm T}  = -\frac{\hbar u\gamma_a^4}{16\pi^2a^4}\sum_{j=1}^{\infty}\sum_{\ell=1}^{\infty}\left\{2\nu_{\ell}^{(\text{i})}\vartheta_{j\ell}(0) + \epsilon_{\ell}^{(\text{i})}\big[\vartheta_{j\ell}(r) + \vartheta_{j\ell}(-r)\big]\right\}.
\label{FreeEtwoplane}
\end{equation}

The free energy density allows us, for instance, to calculate thermodynamic observables such as the entropy density of the liquid. However, before analyzing the entropy density, let us first investigate the low- and high-temperature limits of Eq. \eqref{FreeEtwoplane}. 

By first performing the sum over \(\ell\) followed by the sum over \(j\), we obtain the leading-order contribution for each boundary condition considered in the low-temperature limit, that is,
\begin{equation}
\mathcal{F}^{\rm 2p}_{\rm T} \simeq
\begin{cases} 
\;\;\;\dfrac{\hbar u}{a^4} \dfrac{\pi^4r^2}{7560}\gamma_a^6 + \mathcal{O}\left(\gamma_a^8\right)  & \text{(Dirichlet)}, \\[0.5em]
-\dfrac{\hbar u}{a^4} \dfrac{\zeta(3)}{16\pi}\gamma_a^3 + \mathcal{O}\left(\gamma_a^4\right) & \text{(Neumann)}, \\[0.5em]
\;\;\;\dfrac{\hbar u}{a^4} \dfrac{\pi^4r^2}{7560}\gamma_a^6 + \mathcal{O}\left(\gamma_a^8\right) & \text{(DN)}, \\[0.5em]
\;\;\;\dfrac{\hbar u}{a^4} \dfrac{\pi^2}{720}\gamma_a^4 + \mathcal{O}\left(\gamma_a^6\right) & \text{(ND)}. 
\end{cases}
\label{lowTfree}
\end{equation}
In the expression above, we note that the Dirichlet and DN cases have the same dominant asymptotic behavior at low temperatures.

In contrast, the high-temperature limit of Eq.~\eqref{FreeEtwoplane} is reached by first performing the sum over \(j\) followed by the sum over \(\ell\). The asymptotic behavior, in this case, is governed by
\begin{equation}
\mathcal{F}^{\rm 2p}_{\rm T} \simeq -\frac{k_BT}{64\pi a^3} \sum_{\ell=1}^{\infty} 
\left\{
\frac{2\nu_{\ell}^{(\text{i})}}{\ell^3} + \epsilon_{\ell}^{(\text{i})}\left[\frac{1}{(\ell + r)^3} + \frac{1}{(\ell - r)^3}\right]
\right\} + \text{C.T.}\,,
\label{highTFree}
\end{equation}
where \text{C.T.} denotes constant terms (temperature-independent) representing the next-to-leading-order contributions. The sum over $\ell$ can be expressed in terms of known special functions. However, what makes this result particularly significant is its exact correspondence with the predicted classical behavior, similar to what happens to the total energy density analyzed in the previous subsection. 

For numerical analysis, we can perform the sum over $j$ in Eq. \eqref{FreeEtwoplane} to obtain the following expression:
\begin{equation}
\mathcal{F}^{\rm 2p}_{\rm T}  = -\frac{\hbar u}{64\pi^2a^4}\sum_{\ell=1}^{\infty} 
\left\{2\nu_{\ell}^{(\text{i})}\Lambda_{T}(\ell)
+ \epsilon_{\ell}^{(\text{i})}\big[\Lambda_{T}(\ell + r) + \Lambda_{T}(\ell - r)\big]
\right\},
\label{FreeEtwoplaneplot}
\end{equation}
where the function $\Lambda_{T}(v)$ has been defined in Eq. \eqref{Lambda_func}.
\begin{figure}[h] 
    \centering 
    \includegraphics[width=0.453\textwidth]{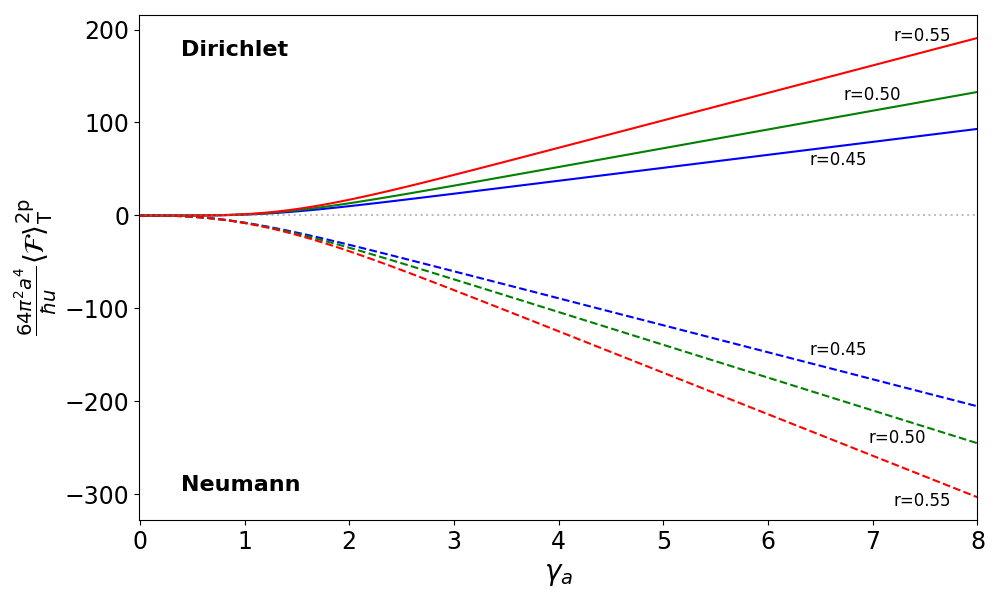}
    \includegraphics[width=0.5\textwidth]{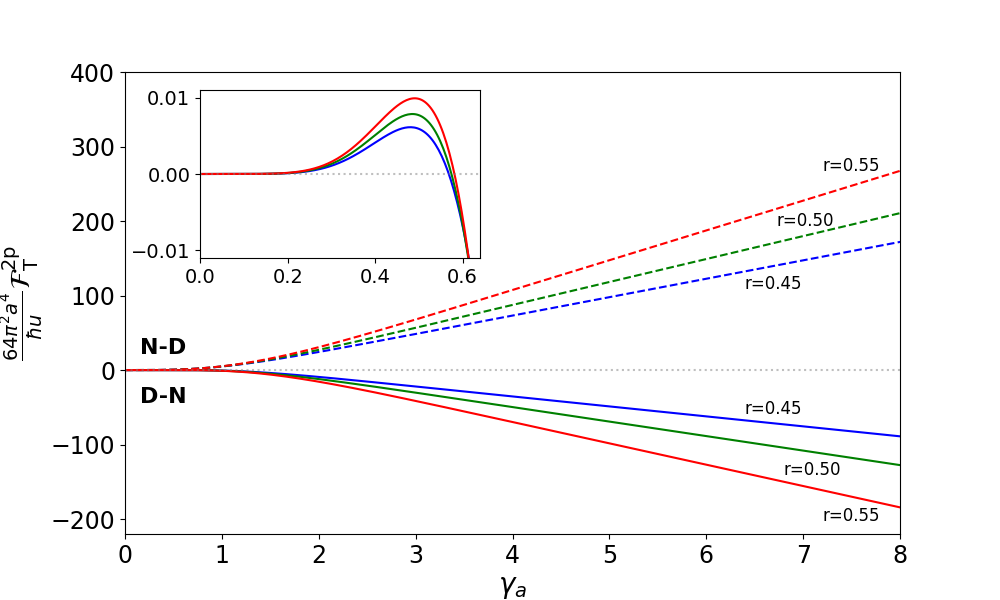}
    \caption{The free energy density from Eq.~\eqref{FreeEtwoplaneplot} is plotted as a function of the dimensionless parameter $\gamma_a$, for various values of $r$ and for each boundary condition case listed in Table~\ref{t1}.}
    \label{fig6} 
\end{figure}

Fig.~\ref{fig6} shows the dimensionless free energy density from Eq.~\eqref{FreeEtwoplaneplot} plotted as a function of the dimensionless parameter $\gamma_a$, for various values of $r$ and for each boundary condition considered. The left side of the figure displays the curves for Dirichlet and Neumann boundary conditions, while the right side presents the results for the DN and ND configurations.

Both plots exhibit the linear temperature dependence predicted by Eq.~\eqref{highTFree} in the high-temperature regime. In contrast, at low temperatures, the Dirichlet boundary condition leads to a curve approaching zero from positive values, while the Neumann case approaches zero from negative values. For the mixed DN and ND cases, both curves approach zero from positive values, as clearly shown in both the main plot and the inset on the right. This behavior is consistent with the analytical expressions given in Eq.~\eqref{lowTfree}.

Now, let us turn to the analysis of the entropy density of the liquid. By using the expression for the free energy density in Eq.~\eqref{FreeE} and the fundamental thermodynamic relation
\begin{equation}
\mathcal{S} = -\frac{\partial \mathcal{F}}{\partial T},
\label{expressionF}
\end{equation}
we obtain the thermal contribution to the entropy density as
\begin{equation}
\mathcal{S}_T = \frac{4k_B}{\pi^2(\hbar u\beta)^3} \sum_{j=1}^{\infty} \sum_{\ell=-\infty}^{\infty} \left\{ \nu_{\ell}^{(\text{i})} \Omega_{j\ell}(0) + \epsilon_{\ell}^{(\text{i})} \Omega_{j\ell}(r) \right\},
\label{entropy0}
\end{equation}
where we have introduced the function
\begin{equation}
\Omega_{j\ell}(r) = \frac{j^2}{\big[j^2 + \gamma_a^2 (\ell + r)^2 \big]^3}.
\label{Omega}
\end{equation}

The Minkowski contribution to the entropy density in the case $\ell=0$ in the first term on the r.h.s. of Eq.~\eqref{entropy0} is given by
\begin{equation}
\mathcal{S}^{\text{M}}_T = \frac{2\pi^2 k_B (k_B T)^3}{45 (\hbar u)^3}.
\label{entropy_M}
\end{equation}
This result is consistent with previous findings in the literature \cite{Bezerra:2011zz}. The expression above represents a purely quantum contribution, as it has no classical counterpart. Moreover, the limit $\hbar \to 0$ cannot be applied, since Planck's constant appears in the denominator of Eq.~\eqref{entropy_M}.

Additionally, the single-plane contribution arises from the second term on the r.h.s. of Eq.~\eqref{entropy0}. This can be expressed as
\begin{equation}
\mathcal{S}^{\text{1p}}_T = \epsilon^{\text{(i)}}_0 \frac{k_B}{32\pi z^3} \left\{ \coth\left(\pi\gamma_z\right) + \pi\gamma_z \left[1 - 2\pi\gamma_z \coth\left(\pi\gamma_z\right)\right] \operatorname{csch}^2\left(\pi\gamma_z\right) \right\}.
\label{entropy1p}
\end{equation}
At high temperatures, the first term on the r.h.s. of Eq.~\eqref{entropy1p} approaches unity, while the second and third terms vanish exponentially. Thus, we observe that the leading-order contribution in this regime is temperature-independent, followed by exponentially suppressed corrections. Since this result is independent of Planck's constant, it represents a classical contribution to the entropy density.

At low temperatures, on the other hand, the single-plane contribution to the entropy density yields a first-order approximation of the form
\begin{equation}
\mathcal{S}^{\text{1p}}_T \simeq \epsilon^{\text{(i)}}_0 \frac{2\pi^2 k_B (k_B T)^3}{45 (\hbar u)^3} + \mathcal{O}\left((k_B T)^5\right).
\label{entropy1plowT}
\end{equation}
Thus, we observe that in the limit \( T \to 0 \), the entropy density vanishes, in agreement with the third law of thermodynamics and, consequently, the Nernst heat theorem. Note that if we had not set the constant \( C \) in Eq.~\eqref{FreeE1} to zero, the entropy density in this limit would instead be \( \mathcal{S}^{\text{1p}}_T = C \). Therefore, to ensure that the entropy density satisfies the third law of thermodynamics, we must set \( C = 0 \).
\begin{figure}[h] 
    \centering 
    \includegraphics[width=0.46\textwidth]{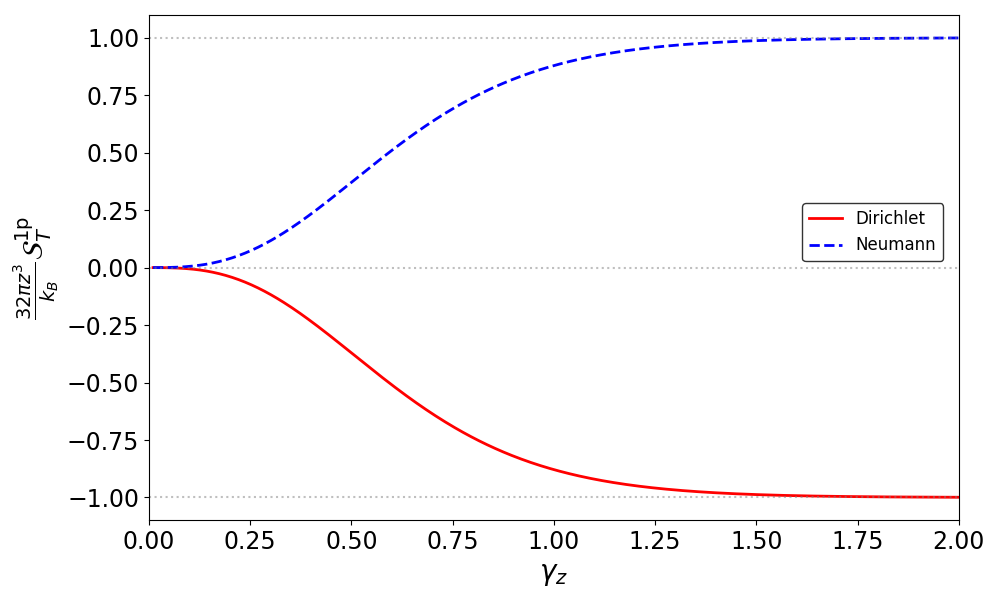}
     \caption{The entropy density from Eq.~\eqref{entropy1p} plotted as a function of $\gamma_z$.}
    \label{fig7} 
\end{figure}

Fig.~\ref{fig7} shows the entropy density curves, calculated using Eq.~\eqref{entropy1p}, as a function of $\gamma_z$ for both Dirichlet and Neumann boundary conditions. The plot demonstrates that the curves approach the constant classical limit at high temperatures and tend to zero at low temperatures, in agreement with the third law of thermodynamics.

The two-plane contribution to the entropy density, given by Eq.~\eqref{entropy0}, can be obtained by subtracting the Minkowski entropy density in Eq.~\eqref{entropy_M} and the single-plane contribution in Eq.~\eqref{entropy1p}. Thus, we obtain
\begin{equation}
\mathcal{S}^{\text{2p}}_T = \frac{k_B \gamma_a^3}{2a^3 \pi^2} \sum_{j=1}^{\infty} \sum_{\ell=1}^{\infty} \left\{ 2\nu_{\ell}^{(\text{i})} \Omega_{j\ell}(0) + \epsilon_{\ell}^{(\text{i})} \bigl[ \Omega_{j\ell}(r) + \Omega_{j\ell}(-r) \bigr] \right\}.
\label{entropy2p}
\end{equation}

As previously done for other observables, we now analyze the two-plane contribution in the low- and high-temperature limits. By first summing over $\ell$ and then over $j$, the low-temperature limit ($\gamma_a \ll 1$) of the above expression exhibits the following asymptotic behavior:
\begin{equation}
\mathcal{S}^{\text{2p}}_T \simeq
\begin{cases} 
\dfrac{k_B}{a^3} \dfrac{3\zeta(3)}{64\pi} \gamma_a^2 + \mathcal{O}\left(\gamma_a^5\right) & \text{(Dirichlet)}, \\[0.5em]
\dfrac{k_B}{a^3} \dfrac{21\zeta(3)}{64\pi} \gamma_a^2 + \mathcal{O}\left(\gamma_a^3\right) & \text{(Neumann)}, \\[0.5em]
-\dfrac{k_B}{a^3} \dfrac{\pi^4 r^2}{630} \gamma_a^5 + \mathcal{O}\left(\gamma_a^7\right) & \text{(DN)}, \\[0.5em]
-\dfrac{k_B}{a^3} \dfrac{\pi^2}{90} \gamma_a^3 + \mathcal{O}\left(\gamma_a^5\right) & \text{(ND)}.
\end{cases}
\label{lowTentropy}
\end{equation}
Once again, it is evident that in the limit $T \to 0$, the entropy vanishes, consistent with the third law of thermodynamics.

Regarding the high-temperature behavior, by first performing the sum over $j$ followed by the sum over $\ell$ as the dominant contribution, we obtain
\begin{equation}
\mathcal{S}^{\text{2p}}_T = \frac{k_B}{32\pi a^3} \sum_{\ell=1}^{\infty} \left\{ \frac{2\nu_{\ell}^{(\text{i})}}{\ell^3} + \epsilon_{\ell}^{(\text{i})} \left[ \frac{1}{(\ell + r)^3} + \frac{1}{(\ell - r)^3} \right] \right\} + \text{E.S.}\,.
\label{entropy2phighT}
\end{equation}
This result is independent of both Planck's constant and temperature, indicating that the leading-order contribution is classical.

A convenient expression for numerical analysis can be obtained by performing the sum over $j$ in Eq.~\eqref{entropy2p} while keeping the sum over $\ell$, which cannot be evaluated analytically. This yields
\begin{equation}
\mathcal{S}^{\text{2p}}_T = \frac{k_B}{32 a^3 \pi} \sum_{\ell=1}^{\infty} \left\{ 2\nu_{\ell}^{(\text{i})} \Delta_T(\ell) + \epsilon_{\ell}^{(\text{i})} \bigl[ \Delta_T(\ell + r) + \Delta_T(\ell - r) \bigr] \right\},
\label{entropy2pplot}
\end{equation}
where we have defined the function
\begin{equation}
\Delta_T(v) = \frac{\coth(\gamma_a\pi v) - \gamma_a\pi v \bigl[ -1 + 2 \gamma_a\pi v \coth(\gamma_a\pi v) \bigr] \operatorname{csch}^2(\gamma_a\pi v)}{v^3}.
\label{Delta}
\end{equation}
\begin{figure}[h] 
    \centering 
    \includegraphics[width=0.47\textwidth]{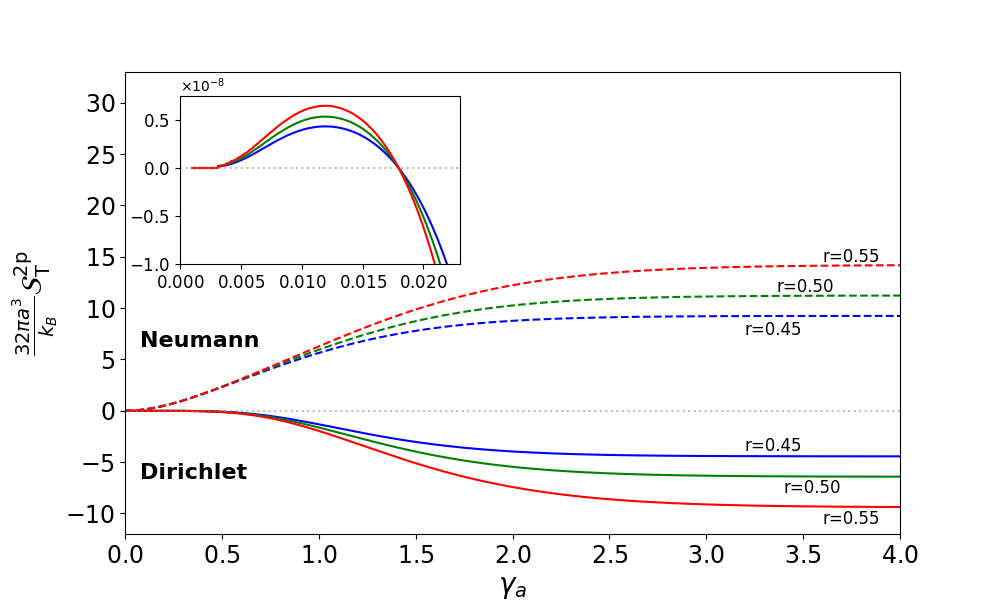}
    \hspace{-0.9cm}
    \includegraphics[width=0.47\textwidth]{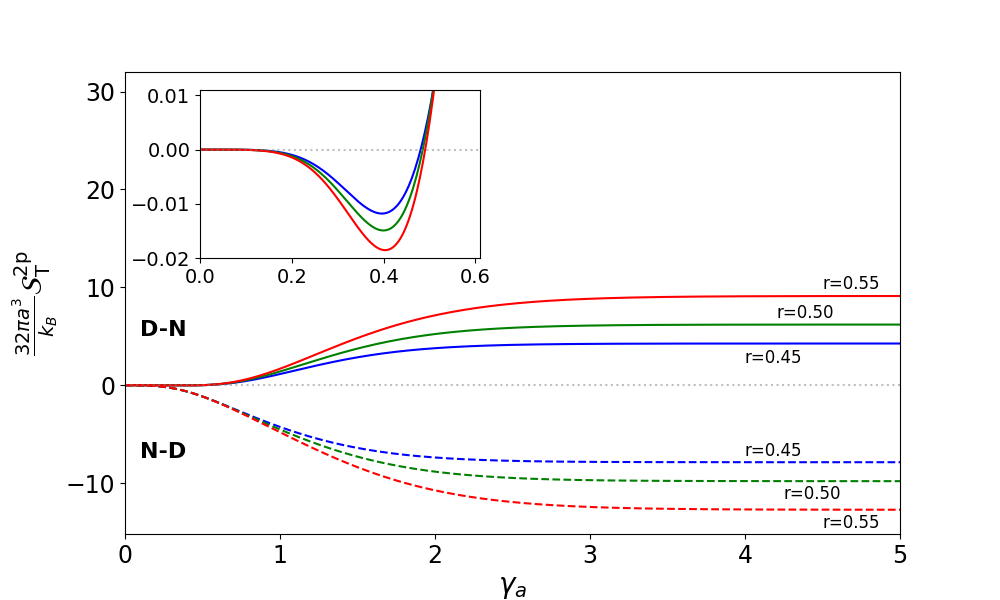}
    \caption{The entropy density from Eq.~\eqref{entropy2pplot} is plotted as a function of the dimensionless parameter $\gamma_a$, for various values of $r$ and for each boundary condition case listed in Table~\ref{t1}.}
    \label{fig8} 
\end{figure}

Fig.~\ref{fig8} presents the entropy density curves computed from Eq.~\eqref{entropy2pplot} as functions of $\gamma_a$ for various $r$ values and for each boundary condition case listed in Table~\ref{t1}. The left plot displays results for Dirichlet and Neumann boundary conditions, while the right plot shows the DN and ND configurations.

Both plots demonstrate that the entropy density approaches a constant value in the classical high-temperature regime, in excellent agreement with Eq.~\eqref{entropy2phighT}. At low temperatures, distinct behavior emerges. The left plot shows both boundary condition cases (Dirichlet and Neumann) vanishing asymptotically from positive values, while the right plot reveals the mixed configurations (DN and ND) approaching zero from negative values. These features are clearly shown in both the main plots and their insets, and are fully consistent with the low-temperature asymptotic behavior predicted by Eq.~\eqref{lowTentropy}.

\section{Conclusions}
\label{sec4}
%
Our investigation of phonon-mediated density fluctuations in confined classical liquids has provided significant theoretical insights with important implications across multiple domains of physics. Through rigorous analytical calculations complemented by numerical analysis, we have systematically characterized how  Dirichlet, Neumann, and mixed (DN/ND) boundary conditions influence thermodynamic quantities across all temperature regimes.

The key outcomes of this work include the derivation of exact closed-form expressions for fundamental quantities including mean square mass density fluctuations, total energy density, and thermodynamic quantities such as free energy and entropy densities for all considered boundary conditions. Our analysis revealed three distinct temperature regimes with characteristic behaviors. A quantum-dominated regime ($\gamma_a \ll 1$) where fluctuations follow power-law scaling and a classical regime ($\gamma_a \gg 1$) exhibiting linear temperature dependence for most quantities. However, some exceptions emerge. the mean square mass density fluctuation vanishes in this regime, while the entropy density approaches a constant value. Importantly, for the mean square mass density, we must additionally impose $\hbar\to 0$ to properly recover the classical limit. Note that there also exists an intermediate crossover region where both quantum and classical effects contribute significantly, displaying rich behavior that interpolates between the limiting cases. This regime may be particularly relevant for experimental observation, as it combines measurable thermal effects with distinctive signatures.

The boundary conditions were shown to imprint characteristic signatures on all physical observables. Particularly noteworthy are the different scaling behaviors in the quantum regime. For the entropy density, for instance, this behavior leads to its consistent vanishing as $T \to 0$ for all boundary configurations, providing strong validation of the third law of thermodynamics within our framework.

In summary, the theoretical implications of these results are of great interest since they establish concrete connections between quantum field theory and hydrodynamics. The demonstrated analogy between phonon fluctuations in confined liquids and quantum vacuum fluctuations extends the conceptual bridge between these domains, with the sound velocity $u$ playing a role analogous to the light velocity $c$ in conventional quantum field theory. While the predicted acoustic Casimir effects are weaker than their electromagnetic counterparts by $\mathcal{O}(10^{-6})$, they nevertheless represent a potentially measurable manifestation of quantum hydrodynamics.

Moreover, although our results draw formal inspiration from quantum field theory, their applicability is directed toward the study of confined fluid systems, classical or semiclassical in nature. The quantized phonon field serves as an effective analog model, with clear physical interpretation rooted in fluid dynamics and statistical mechanics. Thus, we do not claim to capture the full complexity of relativistic vacuum physics. 

{\acknowledgments}
H.M is partially supported by the National Council for Scientific and Technological Development (CNPq) under grant
No 308049/2023-3. K.E.L.F would like to thank the Paraíba State Research Foundation (FAPESQ) for financial support. 
\\


\end{document}